\documentclass[preprint,superscriptaddress,preprintnumbers,aps,amsmath,amssymb]{revtex4}
\usepackage{bm}
\usepackage{color}
\usepackage[pdftex]{graphics}

\usepackage{graphicx}%

\usepackage{multirow}
\usepackage{setspace}
\usepackage{array}
\usepackage{amsmath}
\usepackage{booktabs}
\usepackage{tabularx}
\usepackage{enumerate}
\usepackage{dcolumn}
\usepackage{mathdots}

\usepackage{amsfonts}
\usepackage{float}
\usepackage{subfigure}
\def\prb#1#2#3{{\it Phys.~Rev.}~B~{\bf #1},\ #2\ (#3)}

\def\jcp#1#2#3{J.~Chem.~Phys.~{\bf #1},\ #2\ (#3)}

\def\pra#1#2#3{{\it Phys.~Rev.}~A~{\bf #1},\ #2\ (#3)}
\def\pre#1#2#3{{\it Phys.~Rev.}~E~{\bf #1},\ #2\ (#3)}
\def\prl#1#2#3{{\it Phys.~Rev.~Lett.}~{\bf #1},\ #2\ (#3)}

\def\k1{k_1}
\def\k2{k_2}
\def\q1{q_1}
\def\q2{q_2}

\def\({\left (}
\def\){\right )}
\def\[{\left [}
\def\]{\right ]}

\newcommand{\beq}{\begin{equation}}
\newcommand{\eeq}{\end{equation}}

\newcommand{\ket}[1]{| #1 \rangle}

\newcount\colveccount
\newcommand*\colvec[1]{
        \global\colveccount#1
        \begin{pmatrix}
        \colvecnext
}
\def\colvecnext#1{
        #1
        \global\advance\colveccount-1
        \ifnum\colveccount>0
                \\
                \expandafter\colvecnext
        \else
                \end{pmatrix}
        \fi
}

\begin{document}
\date{\today}
\title{Quantum walk and Anderson localization of rotational excitations in disordered ensembles of polar molecules
}  
\author{T. Xu and R. V. Krems}
\affiliation{Department of Chemistry, University of British Columbia, Vancouver, BC V6T 1Z1, Canada}

\begin{abstract}
We consider the dynamics of rotational excitations placed on a single molecule in spatially disordered 1D, 2D and 3D ensembles of ultracold molecules trapped in optical lattices. 
      The disorder arises from incomplete populations of optical lattices with molecules. This leads to a model corresponding to a quantum particle with long-range tunnelling amplitudes moving on a lattice with the same on-site energy but with forbidden access to random sites (vacancies).  
 We examine the time and length scales of Anderson localization for this type of disorder with realistic experimental parameters in the Hamiltonian. 
     We show that for an experimentally realized system of KRb molecules on an optical lattice this type of disorder leads to disorder-induced localization in 1D and 2D systems on a time scale $t \sim 1$ sec.     
For 3D lattices with $55$ sites in each dimension and vacancy concentration $ 90~\%$,  the rotational excitations diffuse to the edges of the lattice and show no signature of Anderson localization.  
We examine the role of the long-range tunnelling amplitudes allowing for transfer of rotational excitations between distant lattice sites. Our results show that the long-range tunnelling has little impact on the dynamics in the diffusive regime but affects significantly the localization dynamics in lattices with large concentrations of vacancies, enhancing the width of the localized distributions in 2D lattices by more than a factor of 2.  Our results raise a general question whether quantum particles with long-range tunnelling can undergo quantum localization in 3D lattices with substitutional disorder.

\end{abstract}

\maketitle

\clearpage
\newpage

\section{Introduction}

Scattering of electrons by impurities in disordered crystals leads to  Anderson localization \cite{anderson,review-rmp}, which determines the conductor - insulator transitions in metals \cite{example-3}, quasi-crystals \cite{example-4} and granular metal films \cite{example-5}, and is associated with many interesting phenomena, such as the quantized phases of the integer Hall effect \cite{review-rmp,example-1}.   If written in a second-quantized form, the Hamiltonian of an electron in a disordered lattice can be mapped onto the Hamiltonians for a variety of systems \cite{mapping}. This mapping can be used to study transport properties of electrons in disordered lattices by examining the properties of other quantum particles and quasi-particles. For example, disordered-induced localization has been studied with microwaves in a tubular waveguide \cite{microwaves, microwaves-2}, optical photons in opaque media \cite{photons, photons-2} and ultracold atoms in optical lattices \cite{atoms-1, atoms-2, atoms-3, atoms-4}. 
Despite these studies, there are still many open questions about Anderson localization. For example, the scaling hypothesis providing the relation between quantum localization in systems of different dimensionality \cite{scaling-hypothesis} still awaits experimental confirmation. Equally important and widely debated are the effects of inter-particle interactions \cite{correlations} and dissipative forces \cite{dissipation-induced-localization} on localization in far-from-equilibrium systems  or the role of long-range tunnelling amplitudes \cite{long-range-prl, long-range-dispersion, long-range-effects} allowing particles to transition between distant lattice sites. 

The development of experimental methods for producing molecules at ultracold temperatures has created new possibilities for studying disorder-induced localization. As demonstrated in recent experiments \cite{jun-ye-0, jun-ye-1,jun-ye-2,jun-ye-nature,hazzard}, an ensemble of ultracold KRb diatoms, produced by photoassociation of ultracold K and Rb atoms and trapped in an optical lattice, forms a random spatial distribution of KRb molecules in the ro-vibrational ground state. If the trapping field of the optical lattice has high intensity, the molecular motion is strongly confined and the spatial distribution is quasi-stationary, eliminating molecule - molecule collisions. The excitation of molecules from the rotational ground state to the lowest rotationally excited state leads to the formation of delocalized excitations \cite{jun-ye-nature,hazzard}, representing the wave packets of rotational Frenkel excitons \cite{our-exciton-paper2, ping-njp-paper}.  The purpose of the present work is to explore if these wave packets can be used as probe particles to study dynamics of localization induced by the disorder potential stemming from the randomness of the molecular positions.

The rotational excitations in an ensemble of polar molecules randomly distributed on an optical lattice possess several unique properties:
\begin{itemize}

\item
 The rotationally excited states of molecules have long radiative lifetimes ($> 1- 10$ sec), allowing one to probe coherent dynamics over long time scales.  

\item
 The rotational excitations can be placed on molecules in well defined lattice sites. 
 This can be achieved by applying a spatial gradient of an electric field and a microwave field pulse resonant with rotational transitions for molecules in the desired part of the lattice \cite{DeMillePRL02}. The rotational excitations thus produced form spatially localized wave packets \cite{ping-njp-paper}. Once the gradient of the field is removed, these wave packets can travel throughout lattice due to resonant energy transfer between molecules, undergoing quantum walk
 \cite{quantum-walk-1,quantum-walk-2}
   and scattering by empty lattice sites. By shaping the gradient of the field, it may be possible to prepare the excitation wave packets representing superpositions of excitations in spatially separated lattice sites, thus enabling the controlled study of the role of initial conditions and entanglement between spatially separated particles on the dynamics of quantum walk.  

\item
The interactions between the rotational excitations can be tuned by applying a DC electric field \cite{biexcitons}.

\item The transfer of rotational excitations between molecules is enabled by long-range dipole - dipole interactions, leading to long-range transfer of excitations between distant lattice sites. This can be used to study the effects of long-range tunnelling on disorder-induced localization. 

\item The translational motion of molecules can be controlled by varying the strength of the optical lattice potential. As shown in Ref. \cite{our-polaron-paper1, our-polaron-paper2}, the translational motion of molecules can be coupled to the rotational excitations. This can be used to study disorder-induced localization and diffusion of quantum particles in the presence of controlled coupling to a dissipative environment. 

\item The optical lattice potential can be designed to produce a three-dimensional (3D) finite lattice of molecules or to separate molecules in one or two dimensions, effectively confining the rotational excitations to  one-dimensional (1D) or two-dimensional (2D) disordered lattices \cite{optical-lattices}. 


\end{itemize}
These properties make the rotational excitations in spatially disordered ensembles of ultracold molecules a unique new platform for the study of quantum walk and disorder-induced localization.  In particular, the ability to change the dimensionality of the optical lattice potential opens up a unique possibility to verify the universality of Anderson localization in different dimensionalities \cite{scaling-hypothesis} and the ability to tune the interactions between the rotational excitations \cite{our-exciton-paper2} suggests a new way to study the effects of particle interactions on Anderson localization. However, the microscopic Hamiltonian describing rotational excitations in an optical lattice 
leads to a specific model corresponding to a quantum particle moving on a lattice with the same on-site energy but with forbidden access to random sites, with the tunnelling amplitudes decaying as the inverse cube of the hopping distance.   
Whether or not such a model can be used to observe Anderson localization in finite, or even infinite \cite{note, quote}, 2D systems and whether or not such a model allows Anderson localization in 3D systems of any size \cite{3D-anderson} is -- to the best of our knowledge -- unknown \cite{3D-note}.




The specific goal of this work is to consider the following questions:

\begin{itemize}

\item[(i)] While quantum particles placed in a 2D disordered potential are generally  expected to undergo Anderson localization \cite{scaling-hypothesis,note,quote}, the localization length is sensitive to the microscopic details of the particle - disorder interactions and can be very large. Given that the experiments with optical lattices employ finite ensembles of molecules, typically with less than 100 sites per dimension, is the disorder potential for rotational excitations in a partially populated optical lattice strong enough to allow for observations of Anderson localization in 2D systems on the length scale of 100 lattice sites? 
  
  \item[(ii)] If the disorder potential is strong enough to induce Anderson localization in finite 2D arrays, what is the minimum concentration of lattice vacancies leading to localization lengths less than 100 lattice sites? 
  
  \item[(iii)]
 Quantum particles placed in a 3D disordered potential may or may not undergo Anderson localization \cite{scaling-hypothesis}, depending on the microscopic details of the particle - disorder interactions. Is the disorder potential for rotational excitations in 3D disordered ensembles strong enough to induce Anderson localization?

\item[(iv)]
What is the effect of the long-range tunnelling amplitudes on the dynamics in the disordered system of this specific kind? 

\item[(v)]
What is the time scale of the disorder-induced localization, if any,  for excitations placed in specific lattice sites? 

  If the disorder potential is too weak, the localization dynamics may not be observable over experimentally feasible time scales. If the disorder potential is too strong, localized rotational wave packets may remain pinned to their original location and not exhibit any dynamics of quantum walk over experimentally feasible time scales. 

\end{itemize}
Questions (i) - (iv) are general for the lattice model with tunnelling amplitudes randomly omitted and the results pertaining to question (v) can be renormalized for a particular system by re-scaling the time dependence of observables. To answer these questions, we consider the dynamics of rotational excitations placed in individual sites of an optical lattice partially populated with KRb molecules. To simulate an experiment with destructive measurements, we average the dynamics of rotational energy transfer between molecules in the disordered ensembles over multiple disorder orientations and examine the time scales of the formation of disorder-induced distributions of rotational excitations in lattices of various dimensionality. We examine the effects of the long-range tunnelling amplitudes in lattices of various dimensionality and discuss the effect of disorder on diffusion of rotational excitations in lattices of various dimensionality.

\section{Methodology}

The Hamiltonian for an ensemble of molecules on an optical lattice of arbitrary dimensionality can be written as 
\begin{eqnarray}
\hat{H} = \sum_{\bm n} \hat H_{\bm n} + \sum_{\bm n} \sum_{\bm n'} \hat V_{\bm n, \bm n'}, 
\end{eqnarray}
where $\hat H_{\bm n}$ is the Hamiltonian of an isolated molecule placed in lattice site
$\bm n$ positioned at $\bm r = \bm n a$, 
$a$ is the lattice constant, and $\hat V_{\bm n, \bm n'}$ is the dipole - dipole interaction between molecules in sites $\bm n$ and $\bm n'$. The vector index is $\bm n = n_x$ for 1D lattices, $\bm n = (n_x, n_y)$ for 2D lattices and $\bm n = (n_x, n_y, n_z)$ for 3D lattices, with each of $n_x$, $n_y$ and $n_z$ running from $- N$ to $ N$, so that ${\cal N} = 2{ N}+1$ is the number of lattice sites per dimension. We assume that the translational motion of molecules can be neglected. This approximation is valid for the trapping field potential with the trapping frequency exceeding $100$ kHz \cite{our-polaron-paper1}. 

Following the experimental work in Ref. \cite{jun-ye-nature}, 
we assume that each molecule is initially in the lowest eigenstate of $\hat H_{\bm n}$, corresponding to zero rotational angular momentum $j=0$ of the molecule. Some of the molecules are subsequently excited to an isolated hyperfine state within the triplet of the hyperfine states corresponding to the rotational angular momentum $j=1$ and the projection of $\bm j$ on the space-fixed quantization axis $m_j = -1$. The lowest energy state and this specific excited state for the molecule in site $\bm n$ are hereafter denoted as $|g_{\bm n}\rangle$ and $|e_{\bm n} \rangle$. The states $|g \rangle$ and $|e\rangle$ have the opposite parity.

Since the dipole - dipole interaction operator can be written as a sum over products of rank-1 spherical tensors  acting on the subspace of the individual molecules, the matrix elements $\langle g_{\bm n}| \langle g_{\bm n'} | \hat V_{\bm n, \bm n'} |g_{\bm n}\rangle |g_{\bm n'}\rangle$ vanish so the molecules are non-interacting when in the ground state. However, the dipole - dipole interaction has non-zero matrix elements between the product states $|g_{\bm n}\rangle |e_{\bm n'}\rangle$ and $|e_{\bm n}\rangle |g_{\bm n'}\rangle$. The matrix elements 
\begin{eqnarray}
t_{\bm n, \bm n'} = \langle g_{\bm n}| \langle e_{\bm n'} | \hat V_{\bm n, \bm n'} |e_{\bm n}\rangle |g_{\bm n'}\rangle
\label{t-elements}
\end{eqnarray}
stimulate resonant energy transfer of the $|g\rangle \rightarrow |e\rangle$ excitation between molecules in sites $\bm n$ and $\bm n'$. For the specific states $|g \rangle = |j = 0 \rangle$ and $|e \rangle = |j = 1, m_j = -1 \rangle$, the matrix elements (\ref{t-elements}) have the following form: 
 \begin{equation}
       t_{\bm n, \bm n'} =\frac{1}{4\pi\epsilon_0}\cdot\frac{d_{\bm{n}}d_{\bm{n'}}}{6a^3|{{\bm{n}}-{\bm{n'}}}|^3}(3\cos^2\theta_{\bm{n}\bm{n'}}-1)
        \label{eq:vdd}
 \end{equation}
where $d_{\bm{n}}=d_{\bm{n'}}=0.57$ Debye is the permanent dipole moment of KRb molecules \cite{jun-ye-nature}, $\theta_{\bm{n}\bm{n'}}$ is the angle between the intermolecular axis joining the molecules in lattice sites $\bm n$ and $\bm n'$ and the $z$-axis. We assume that $\theta_{\bm{n}\bm{n'}}=\pi/2$ for 1D and 2D lattices. The molecules are trapped in a lattice with lattice constant $a=532$ nm \cite{jun-ye-nature}. For the specific system considered here, the value of these matrix elements is 
\begin{eqnarray}
 t_{\bm n, \bm n'} =\alpha \times \frac{(3\cos^2\theta_{\bm{n}\bm{n'}}-1)}{{{|\bm{n}}-{\bm{n'}}}|^3}.
\label{alpha-eq}
\end{eqnarray}
with $\alpha = 52.12$ Hz.

The matrix elements (\ref{t-elements}) also lead to the delocalized character of the rotational excitations. In general, the $|g\rangle \rightarrow |e\rangle$ excitation generates the following many-body excited state:
\begin{eqnarray}
|\psi_{\rm exc}\rangle = \sum_{\bm n} C_n |e_{\bm n}\rangle \prod_{{\bm i}
\neq {\bm n}} |g_{\bm i}\rangle . \label{basis}
\end{eqnarray}
In an ideal lattice, the coefficients  $C_{\bm n} =
e^{ia{\bm p} \cdot {\bm n}}/\sqrt{\cal N}$ and  $|\psi_{\rm exc} \rangle \Rightarrow
|\psi_{\rm exc}(\bm p)\rangle$ represents a rotational
Frenkel exciton
with wave vector $\bm p$ \cite{agranovich} , i.e. a rotational excitation completely delocalized over the entire lattice. 
If the lattice is disordered, the excited state is a localized coherent superposition of the Frenkel excitons with different $\bm p$ \cite{ping-njp-paper}. 

When the lattice is partially filled with molecules, the excitation energy of the molecular ensemble depends on the location of molecules in the excited state. 
The excited states of the many-body system of molecules in a disordered lattice are the eigenstates of the following Hamiltonian: 
\begin{eqnarray}
\hat{H} = \sum_{\bm n} w_{\bm n} c^\dagger_{\bm n} c_{\bm n} + \sum_{\bm n} \sum_{\bm n'} t_{\bm n, \bm n'} c^\dagger_{\bm n} c_{\bm n'},
\label{hamiltonian}
\end{eqnarray}
where the operator $c_{\bm n}$ removes an excitation from site $\bm n$, $w_{\bm n}$ is the excitation energy of a molecule in site $\bm n$, if the site is populated, and zero otherwise.  
The transition amplitudes $t_{\bm n, \bm n'}$ are given by Eq. (\ref{t-elements}) if both of the sites $\bm n$ and $\bm n'$ are populated by molecules and zero otherwise. It is this disorder in the transition amplitudes $t_{\bm n, \bm n'}$ that localizes the delocalized excitations. 

In order to compute the time-evolution of the rotational excitations, we diagonalize the matrix of the Hamiltonian (\ref{hamiltonian}) in the site-representation basis $|\bm{n} \rangle \equiv c^\dagger_{\bm n} | 0\rangle$, where the vacuum state is
\begin{eqnarray}
|0\rangle = \prod_{\bm m} |g_{\bm m} \rangle,
\end{eqnarray}
with both $\bm m$ and $\bm n$ running only over the lattice sites populated with molecules. 

The eigenstates of the full Hamiltonian (\ref{hamiltonian}), 
      \begin{equation}
        \ket{\lambda}=\sum\limits_{\bm{n}} C_{\bm{n}}^\lambda\ket{\bm{n}},
        \label{eq:es}
      \end{equation}
are then used to compute the time evolution of the excitation wave packet as follows
      \begin{equation}
        \ket{\psi(t)}=\sum\limits_\lambda C_\lambda e^{-iE_\lambda t/\hbar}\ket{\lambda} = \sum_{\bm n}
        f_{\bm n}(t) | \bm{n} \rangle,
                \label{eq:sT}
      \end{equation}
where $E_{\lambda}$ are the eigenvalues of the Hamiltonian (\ref{hamiltonian}), the coefficients 
\begin{eqnarray}
f_{\bm n}(t) = \sum_{\lambda} C_{\bm n}^\lambda C_{\lambda} e^{-iE_\lambda t/\hbar},
\end{eqnarray}      
and the coefficients $C_{\lambda}$ are the projections of the excitation wave packet at $t=0$ onto the eigenstates of the Hamiltonian, 
      \begin{equation}
        C_\lambda={\langle \lambda}|{\psi(t=0)\rangle}.
        \label{eq:s0exp}
      \end{equation}
The quantities $|f_{\bm n}(t)|^2$ are averaged over multiple calculations with different realizations of disorder (i.e. different random distributions of empty lattice sites). 
            

As will be demonstrated in the following section, the rotational excitations scattered by the empty lattice sites form peaked distributions.  In this article, we will use two measures of distributions: the distribution width $L$ and the standard deviation $\sigma_r$. The distribution width $L$ is 
defined as the length of the lattice (for a 1D lattice), the radius of a circle (for a 2D lattice) or the radius of a sphere (for a 3D lattice) containing 90 \% of the excitation probability.  The standard deviation is defined as $\sigma_r = \sqrt{\langle r^2 \rangle - \langle r \rangle^2} $, where $r^2 = x^2$ for a 1D lattice, $r^2 = x^2 + y^2$ for a 2D lattice and $r^2 = x^2 + y^2 + z^2$ for a 3D lattice. The distribution width characterizes the physical spread of the rotational excitation over the molecular ensemble. However, this quantity may be significantly affected by fluctuations due to specific disorder realizations. The standard deviation depends on the shape of the distribution but is less affected by the fluctuations. We will use the standard deviation in order to characterize non-ambiguously the dynamical properties such as the time dependence of the distributions.

\section{Quantum walk of rotational excitations}

As originally proposed by DeMille \cite{DeMillePRL02}, ultracold molecules trapped in an optical lattice can be selectively excited by applying a gradient of a DC electric field that separates the rotational energy levels of molecules in different lattice sites by a different magnitude. This method can, in principle, be used to excite rotationally a single molecule in a specific lattice site. If the gradient of the field is removed, the rotational excitation thus generated may be transferred to molecules in other lattice sites by resonant energy transfer. The dynamics of this energy transfer can be probed by applying another gradient of the DC electric field at time $t$ and detecting either the excited molecules or molecules remaining in the ground state using resonantly-enhanced multiphoton ionization (REMPI). The REMPI detection is a destructive measurement. If the optical lattice is partially populated with randomly distributed molecules, 
repeating the experiment multiple times to determine the dependence of rotational energy transfer on $t$ is equivalent to averaging over different realizations of disorder (i.e. different distributions of empty lattice sites).

 In order to simulate the outcome of such measurements, we consider an optical lattice partially populated with randomly distributed KRb molecules. We compute the time evolution of a rotational excitation placed in a single lattice site at $t=0$ and average the results at each $t$ over multiple random distributions of empty lattice sites with a fixed concentration of vacancies. Figures 1, 2 and 3 illustrate the characteristic distributions of rotational excitations as a result of quantum walk of a single rotational excitation initially placed in the centre of a 1D, 2D (square) or 3D (cubic) lattice, respectively. In the following sections, we explore the detailed dynamics of how these distributions are formed and the role of long-range tunnelling effects in determining the quantum walk dynamics.

 \subsection{Localization dynamics in 1D lattice}
 
A quantum particle placed in a 1D disordered lattice must undergo Anderson localization \cite{anderson}.  Mapped onto a rotational excitation travelling in a disordered ensemble of molecules,
this well-known result is reflected in the formation of the exponentially localized distributions shown in Figure 1. These distributions characterize the probability of molecules in the corresponding lattice sites to be in the rotationally excited state. 
Figure 1 illustrates that rotational excitations in an optical lattice partially populated with molecules behave as quantum particles in a disordered lattice. It is important to note that the distributions presented in Figure 1 are time-independent. 
 Although the rotational excitation initially placed in a specific lattice site represents a wave packet that, given sufficient time, can potentially explore the entire lattice and exhibit revivals, the averaging over disorder realizations leads to stationary distributions in the limit of long time. In order to illustrate this, we present in Figure 4 the distributions of the rotational excitation computed at different times.

 The lower panel of Figure 4 illustrates the time dependence of the distribution width $L$ in a 1D lattice with the vacancy concentrations 10 \% and 20 \%.  The results show that for the parameters representing rotational excitations in 
 an ensemble of KRb molecules, the averaged excitation probability distributions change most significantly at $t$ between zero and 100 - 300 ms. The distribution widths exhibit a temporary peak at short times. This oscillation survives the averaging over disorder realizations and is present in the results for different concentrations of vacancies. Since the distribution width presented in Figure 4 is much smaller than the lattice size ($1001 \times a$), this short-time oscillation cannot be due to reflection from the lattice boundaries and is likely the result of constructive interference due to back-scattering of the rotational excitation from the empty lattice sites. 
  
As indicated by the results in the lower panel of Figure 4,  the approach to the stationary distributions is determined by the strength of the disorder potential, i.e. the concentration of lattice vacancies. This is illustrated more clearly in the upper panel of Figure 5 showing the time dependence of the standard deviations $\sigma_r$ for different concentrations of vacancies. Figure 5 demonstrates that the rotational excitations in a 1D lattice of KRb molecules reach the time-independent distributions in less than one second, provided the vacancy concentration is $> 3 \%$. 

It is instructive to examine the dependence of the minimum time required to form the stationary distributions on the concentration of vacancies. Figure 6 shows the time $\tau$ required for a rotational excitation initially placed in the center of a lattice to form a time-independent distribution as a function of vacancy concentration in lattices of different dimensionality. Interestingly, Figure 6 reveals that the dependence of $\tau$ on the vacancy concentration is qualitatively  different for lattices of different dimensionality. In 1D lattices, $\tau$ is a monotonically decaying function of the vacancy concentration. This illustrates that increasing the disorder of 1D lattices impedes the dynamics of quantum walk leading to narrower distributions of the rotational excitation (cf., Figure 1) and decreases the time required to reach the stationary distributions (less time since the quantum walker has less space to explore).  
  The calculations shown in Figure 6 have two implications: (i) the time evolution of the rotational excitations in 1D disordered lattices can be used as a probe of the disorder strength; (ii) the mechanisms for the formation of the distributions shown in Figures 1 - 3 are different for lattices of different dimensionality.
In the following section, we correlate the rising or falling dependence of $\tau$ on the concentration of vacancies with the regime of classical diffusion or sub-diffusion, arguing that the sign of the gradient of $\tau$ can be used as a signature of sub-diffusion stimulated by Anderson localization.

\subsection{Localization and diffusion in 2D and 3D lattices}

As can be seen from Figures 2 and 3 and the comparison of the different curves in Figures 5 and 6, the dynamics of rotational energy transfer between molecules trapped in 2D and 3D lattices is quantitatively and qualitatively different from the localization dynamics in 1D lattices. It is apparent that the formation of localized distributions requires higher concentrations of vacancies and the approach to time-independent distributions takes longer time in lattices of higher dimensionality. 
In particular, Figure 6 reveals that the dependence of  the time $\tau$ required to reach a stationary distribution on the vacancy concentration is qualitatively  different for lattices of different dimensionality. In 1D lattices, $\tau$ is a monotonically decaying function of the vacancy concentration. In 2D lattices, $\tau$ exhibits a maximum as a function of the vacancy concentration. In 3D lattices, $\tau$ increases monotonically with the disorder strength. 

In order to elucidate the dynamics of quantum walk in higher dimensions and the implications of Figure 6, we present in Figure 7 the time dependence of the distribution widths $L$ computed for a rotational excitation in 2D and 3D lattices of varying size with varying concentrations of vacant sites. 
The results for the 2D lattices illustrate that at low concentration of vacancies (20 \%) corresponding to the range, where the dependence of $\tau$ on the disorder strength rises, the rotational excitation width approaches the size of the lattice in the limit of long time. This shows that the localization length of the excitation for this concentration of vacancies is greater than the size of the lattice. When the concentration of empty sites is 70 \%, the rotational excitation in a 2D lattice is well localized and the different curves shown in Figure 7 converge to the same value in the limit of long time. This localization regime corresponds to the range in Figure 6, where $\tau$ decreases with the disorder strength. 

Since the experiments with ultracold molecules on optical lattices employ finite samples of molecules, typically $\sim 100$ per dimension \cite{jun-ye-nature}, it is important to find the threshold concentration of vacancies that leads to localized states in 2D lattices with dimension $< 100 \times 100$. By repeating the calculation of Figure 7 for different concentrations of vacancies, we find that $L(t \rightarrow  \infty) < 100$, when the concentration of vacancies exceeds $50 \%$.  This is consistent with the results in Figure 6, which shows that the dependence of $\tau$ on the vacancy concentration exhibits a maximum at the concentration of vacancies between $40 \%$ and $50 \%$. We conclude that the rotational excitations can be localized in 2D disordered lattices of molecules with $100 \times 100$ lattice sites, provided the concentration of vacancies is $> 50 \%$.

The calculations for the rotational excitation dynamics in 3D lattices (lower panel of Figure 7) show that, 
 for each of the lattices considered, the distribution width approaches the size of the lattice in the limit of long time, even for the concentration of empty sites as large as 90 \%. This indicates that the rotational excitation in 3D lattices with 90 \% of empty sites is allowed to diffuse to the edges of the lattice. This is also consistent with the results in Figure 6 showing that $\tau$ rises monotonically for 3D lattices. This indicates that the rotational excitations do not exhibit Anderson localization on the length scale considered. To prove this, we compare the dynamics of the spatial expansion of the rotational excitations in 3D disordered lattices with classical diffusion. 
 
The diffusion of quantum particles in disordered lattices can be examined for 
an alternative signature of Anderson localization.
A quantum particle with cosine dispersion placed in an individual site of an ideal lattice 
expands ballistically to form a distribution with the standard deviation 
 $\sigma_r^2 = \langle r^2 \rangle - \langle r \rangle^2 \propto t^2$ \cite{diffusion,diffusion-2}. In contrast, the diffusion of a classical particle in Brownian motion is 
characterized by $\sigma_r^2  \propto t$. In the presence of Anderson localization, the diffusive dynamics must be suppressed. 

 The time-dependence of $\sigma_r$ can thus be used as a quantitative measure of quantum interference, constructive or destructive, on particle propagation in disordered lattices.

The upper panels of Figure 8 (2D lattices) and Figure 9 (3D lattices) show the time dependence of 
$\sigma_r^2$ computed for 2D and 3D lattices with various concentrations of empty lattice sites. 
As expected, we find that $\sigma_r^2 \propto t^2$ for ideal lattices, both in 2D and 3D. However, the time-dependence of $\sigma_r^2$ is dramatically modified by the presence of vacancies. In particular, Figure 8 illustrates that the rotational excitation in disordered lattices expands ballistically at short times with $\sigma_r^2 \propto t^2$ and exhibits the dependence $\sigma_r^2 \propto t$, characteristic of the diffusive regime, at later times. That $\sigma_r^2$ becomes a linear function of time illustrates that the presence of vacancies destroys interferences accelerating diffusion of quantum particles \cite{motional-narrowing, chaos-diffusion-1, chaos-diffusion-2,vardi}.
At even later times ($t > 300$ ms), the dependence of  $\sigma_r^2$ on time departs from linearity and 
the propagation of the rotational excitations enters a sub-diffusive regime ($\sigma_r^2 \propto t^\gamma$ with $\gamma < 1$). This can happen either because of Anderson localization or because of the reflection from the boundaries.



Since the calculations of Figures 8 and 9 are performed for finite lattices, it is important to examine the effects of the boundaries. In particular, it is important to determine if the departure of  $\sigma_r^2$ from linearity is due to disorder or due to the finite size effects. In order to do this, we first consider the ballistic expansion of the rotational excitations in ideal lattices of varying size (upper right panels of Figures 8 and 9). We observe that, for each lattice size, $\sigma_r^2$ increases as a quadratic function of time at $t < t_1$, until the reflection from the boundaries slows down the ballistic expansion.  We denote by $t_1$ the time, at which the dependence of $\sigma_r^2$ on time in an ideal lattice departs from the quadratic function due to the reflection from the boundaries. The insets of Figures 8 and 9 show the dependence of $t_1$ on the lattice size.   In the presence of disorder, the expansion is significantly slower and the 
reflection from the boundaries must occur at times later than $t_1$. Therefore, we can use 
the time $t_1$ as the lower limit for the onset of the boundary effects and conclude that the dynamics, whether in disordered or ideal lattices, is not affected by the boundary effects at $t < t_1$. 

The lower panel of Figure 8 shows that the dynamics of rotational excitations in 2D lattices with the disorder concentration $> 50 \%$ enters the sub-diffusive regime at times $t < t_1$. 
This  proves that the transition to the sub-diffusive regimes at these concentration of vacancies is due to disorder-induced localization, and not due to the boundary effects. 
For the disorder concentration $20 \%$, we observe that the departure of $\sigma_r^2$ from linearity occurs at $t > t_1$. 
These observations are consistent with the results of Figures 6 and 7. 
We thus conclude that the increase of the time required for the formation of the stationary distributions shown in Figure 6 reflects that increasing the disorder strength slows down the diffusion of rotational excitations to the lattice boundaries, hence the dramatically different dependence of $\tau$ on the disorder strength for 2D (at low vacancy concentrations) and 3D lattices from that in 1D lattices.

  For 3D lattices, the transition to the sub-diffusive regimes always occurs at $t > t_1$,
indicating (but not proving) that the departure of $\sigma_r^2$  from linearity is due to the finite size of the system, and not the scattering by disorder. The lower panel of Figure 9 shows the results for an extreme case of a 3D lattice with 90 \% of the lattice sites vacant.  To prove that the disorder does not lead to the sub-diffusive regime in 3D lattices, we perform the calculations illustrated in the lower panel of Figure 9 for different lattice sizes. For each lattice size, we determine the time $t_2$, at which the expansion of $\sigma^2_r$ undergoes the transition to the sub-diffusive regime. We find that the dependence of $t_2$ on the lattice size is a linear function of the lattice size for extended 3D lattices ($31^3$ -- $55^3$ sites), as expected for a particle in the diffusive regime. This proves that there is no Anderson localization in 3D lattices on the length scale of $55^3$ lattice sites for the disordered system considered here.


\subsection{Effect of long-range tunnelling}

Most models of quantum particles in disordered lattices are based on the nearest neighbour (tight-binding) approximation assuming that 
the tunnelling amplitudes $t_{\bm n, \bm n'}$ in Eq. (\ref{hamiltonian}) are non-zero only when $\bm n$ and $\bm n'$ are adjacent lattice sites. 
However, the tunnelling amplitudes (\ref{t-elements}) responsible for rotational energy transfer between molecules decay as $\propto 1/|\bm n - \bm n'|^3$ functions of the lattice site separation. 
This makes the rotational excitations in an ensemble of polar molecules a new platform for the experimental investigation of long-range tunnelling effects on the formation of localized states. 
The $1/|\bm n - \bm n'|^3$ dependence of the tunnelling amplitudes is particularly interesting
\cite{long-range-prl}. Beyond allowing for direct transitions between distant lattice sites, it leads to non-analyticity of particle dispersion at low particle velocities \cite{long-range-dispersion}. It may also lead to interesting long-range correlations in the localization dynamics of multiple interacting particles \cite{dipolar, long-range-correlations}. 

The effect of the long-range character of the tunnelling amplitudes $t_{\bm n, \bm n'}$ can be -- in principle -- investigated by comparing the dynamics of rotational excitations in ensembles of polar molecules with those in ensembles of homonuclear diatomic molecules.  Ultracold homonuclear molecules have been produced in multiple experiments worldwide \cite{njp-review}. The rotational excitations in an ensemble of non-polar molecules can be resonantly transferred between molecules due to quadrupole - quadrupole interactions, the leading term in the multipole expansion of the interaction between non-polar molecules separated  by a large distance. The quadrupole - quadrupole interaction decays as a $\propto 1/R^5$ function of the molecule - molecule separation $R$. Therefore, the rotational excitations in an ensemble of homonuclear molecules trapped in an optical lattice should be expected to undergo quantum walks governed by the Hamiltonian (\ref{hamiltonian}) with 
$t_{\bm n, \bm n'} \propto 1/|\bm n - \bm n'|^5$. 

In this section, we compare the localization dynamics of particles with the tunnelling amplitudes $t_{\bm n, \bm n'} \propto 1/|\bm n - \bm n'|^3$ and $t_{\bm n, \bm n'} \propto 1/|\bm n - \bm n'|^5$. To remove ambiguity associated with the different magnitudes of the dipole and quadrupole moments, we compare the dynamics of rotational excitations stimulated by the tunnelling amplitudes as given in Eq. (\ref{t-elements}) and the tunnelling amplitudes $\tilde t_{\bm n, \bm n'} = b/|\bm n - \bm n'|^5$, where the constant $b$ is chosen such that $\tilde t_{\bm n, \bm n'}  = t_{\bm n, \bm n'} $ for $|\bm n - \bm n'| = 1$. In other words, we enhance the quadrupole - quadrupole interaction to be the same as the dipole - dipole interaction between molecules in adjacent lattice sites so that the difference between the calculations with $\tilde t_{\bm n, \bm n'} $ and $t_{\bm n, \bm n'} $ arises solely from the different radial dependence of the tunnelling amplitudes. 

The results presented in Figure 10 illustrate that the long-range matrix elements $t_{\bm n, \bm n'}$ lead to less localized distributions in all dimensions. The effect of the long-range matrix elements appears to depend both on the strength of the disorder potential and the dimensionality. To elucidate these dependencies, we present in Figure 11 the ratio of the distribution widths computed in the limit of long time with the long-range amplitudes $t_{\bm n, \bm n'} \propto 1/|\bm n - \bm n'|^3$ and the short-range amplitudes $\tilde t_{\bm n, \bm n'} \propto 1/|\bm n - \bm n'|^5$. The long-range matrix elements appear to have little effect at low concentrations of empty sites, but increase the distribution widths by a factor of 2 or more at large concentrations of vacancies.


The results presented in Figure 11 indicate that the long-range tunnelling amplitudes $t_{\bm n, \bm n'}$ play little role at low disorder strengths in lattices of higher dimensionality, i.e. when the dynamics is expected to be more dominated by diffusion. At the same time, Figure 11 shows that the long-range matrix elements become exceedingly important in the regime where disorder-induced localization is expected to dominate. 
In order to illustrate this more clearly, we show in Figure 12 the distribution widths $L$ computed as functions of time for 2D lattices with two concentrations of vacancies: 20 \% corresponding to the regime dominated by diffusion (cf. Figures 6 and 8) and 70 \% corresponding to the regime, where the rotational excitations in 2D lattices are well localized. The results show that the long-range tunnelling amplitudes insignificantly accelerate the expansion dynamics of the rotational excitation in the diffusive regime, enhancing the distribution widths at short times by about 20 \% or less (upper panel of Figure 12). In stark contrast, the distribution widths obtained with the two types of tunnelling amplitudes are dramatically different in the regime of strong localization (lower panel of Figure 12).


\section{Conclusion}

In this work, we study the dynamics of quantum walk of rotational excitations in finite disordered 1D, 2D and 3D ensembles of ultracold molecules trapped in optical lattices. The disorder arises from incomplete populations of optical lattices with molecules. Thus, the rotational excitations travelling between molecules in a partially populated optical lattice are described by the Hamiltonian (\ref{hamiltonian}), which can be obtained 
from the tight-binding lattice model by randomly omitting some of the tunnelling amplitudes and extending the remaining tunnelling amplitudes  to decay as $\propto \alpha/|n-n'|^3$. While the dynamics of particles described by such a Hamiltonian should be expected to exhibit the general features of quantum particles in a disordered potential, whether or not such a model can be used to observe Anderson localization in finite 2D systems and whether or not such a model can be used to observe Anderson localization in 3D systems (finite or infinite), depends on the microscopic details of particle - disorder interactions. We note that there are only two free parameters in this Hamiltonian: the magnitude of the constant $\alpha$ in Eq. (\ref{alpha-eq}) and the concentration of empty lattice sites. While the magnitude of $\alpha$ determines the time scales of quantum walk and localization dynamics, the localization length in 1D and 2D systems and the presence or absence of Anderson localization in 3D systems is entirely determined by the concentration of vacancies. 

In order to guide the ongoing experiments with ultracold molecules, we chose the value of $\alpha$ to correspond to a specific experimentally realized system of KRb molecules on an optical lattice with lattice constant $a = 532$ nm. We have confirmed that increasing the value of $\alpha$ accelerates the time evolution of the dynamics but leads to the same results in the limit of long time. The qualitative features of the time-dependent results and the results in the limit of long time are, therefore, general for any system described by the Hamiltonian considered here. The time-dependent results can be renormalized for other systems by changing the value of the coupling constant in Eq. (\ref{t-elements}), which controls the time-evolution of the system.

The main results of this work can be summarized as following:
\begin{itemize}

\item For the specific system of 1D arrays of KRb molecules on an optical lattice with lattice constant $a = 532$ nm,  the rotational excitations placed in individual lattice sites form localized distributions within $t \sim 1$ sec. With only 10 $\%$ of vacancies, the width of the exponentially localized distributions is $< 40$ $a$. This indicates that Anderson localization can be studied with rotational excitations in 1D arrays of molecules, providing coherence can be preserved for longer than 1 second.

\item In 2D, this particular type of disorder requires concentrations of empty lattice sites  $ > 50 \%$ in order to  localize the rotational excitations within $100 \times 100$ sites. 


\item
 For 3D disordered arrays with $55$ sites in each dimension, we observe no localization at vacancy concentrations $\leq 90~\%$, i.e. the rotational excitations generated in the middle of the lattice diffuse to the edges of the lattice as classical particles. The vacancy concentration $90~\%$ is near the percolation threshold for a 3D network of sites \cite{jun-ye-nature}.
We note that the results of Ref. \cite{3D-anderson} indicate that the model for the particle with nearest-neighbour hopping on a lattice with random on-site energies and substitutional disorder does allow quantum localization in 3D. Figure 5 of Ref. \cite{3D-anderson} suggests that quantum localization is present in a system with substitutional disorder even in the absence of disorder in the on-site energies. Our results thus indicate (but not yet prove) that the interval of quantum localization in Figure 5 of Ref. \cite{3D-anderson} between the regimes of diffusion and classical localization may be reduced to zero due to long-range tunnelling. This raises a general question whether quantum particles with long-range tunnelling can undergo quantum localization in 3D lattices with substitutional disorder.


\item Our results show that the long-range character of the tunnelling amplitudes in the Hamiltonian (\ref{hamiltonian}) has little impact on the dynamics of particles in the diffusive regime (i.e. in 2D and 3D lattices with low concentrations of vacancies) but affects significantly the localization dynamics in lattices with large concentrations of vacancies, enhancing the width of the localized particle distributions in 2D lattices by a factor of 2.

\item Our results show that the diffusive vs localization regime can be identified by measuring the dependence of the minimum time $\tau$ (cf. Figure 6) required for a particle placed in a specific lattice site to form a time-independent averaged distribution as a function of vacancy concentration.

\item
Our calculations illustrate that the diffusion of rotational excitations placed in 2D lattices with disordered tunnelling amplitudes has three distinct time scales, which can be varied by changing the concentration of empty lattice sites.  At short times, the dynamics is characterized by the quadratic time dependence of $\sigma_r^2 = \langle r^2 \rangle - \langle r \rangle^2$, characteristic of quantum particles with cosine dispersion.  The rise of $\sigma_r^2$ is suppressed by the disorder at later times. This suppression leads to an extended time interval, where $\sigma_r^2$ exhibits a linear dependence on time, characteristic of classical diffusion. This interval of time increases with decreasing disorder strength and can be as long as $\sim100$ ms for the dynamics of in a 2D lattice with 50 $\%$ of empty sites. At still longer times, the dependence of $\sigma_r^2$ on time is sub-diffusive, i.e. slower than linear. This suggests that rotational excitations in 2D disordered ensembles of polar molecules can be used to study the crossover from ballistic expansion to classical diffusion, where quantum interferences enhancing diffusion are suppressed. At longer times, the dynamics of rotational excitations in 2D lattices can be used to study the crossover from classical diffusion to a sub-diffusive regime, where quantum interferences leading to localization become important.

\end{itemize}

In our previous work \cite{our-polaron-paper1}, we showed that the rotational excitations of molecules on an optical lattice can be coupled to the translational motion of molecules, leading to the formation of polarons with a wide range of tunable properties \cite{our-polaron-paper2}. We also showed that, if the molecules on an optical lattice are subjected to DC electric fields, the rotational excitations exhibit strong interactions that may -- depending on the electric field strength -- lead to the formation of quasi-particle bound pairs \cite{biexcitons}. In combination with this previous work, the present results suggest an exciting research avenue for the study of quantum diffusion through disordered lattices in the presence of dissipation and the role of inter-particle interactions on Anderson localization.

\section*{Acknowledgment}
We acknowledge useful discussions with Joshua Cantin, Marina Litinskaya, Evgeny Shapiro, John Sous and Ping Xiang. 
The work was supported by NSERC of Canada.

\clearpage
\newpage

\begin{figure}[ht]
\label{figure1}
\begin{center}
\includegraphics[scale=0.7]{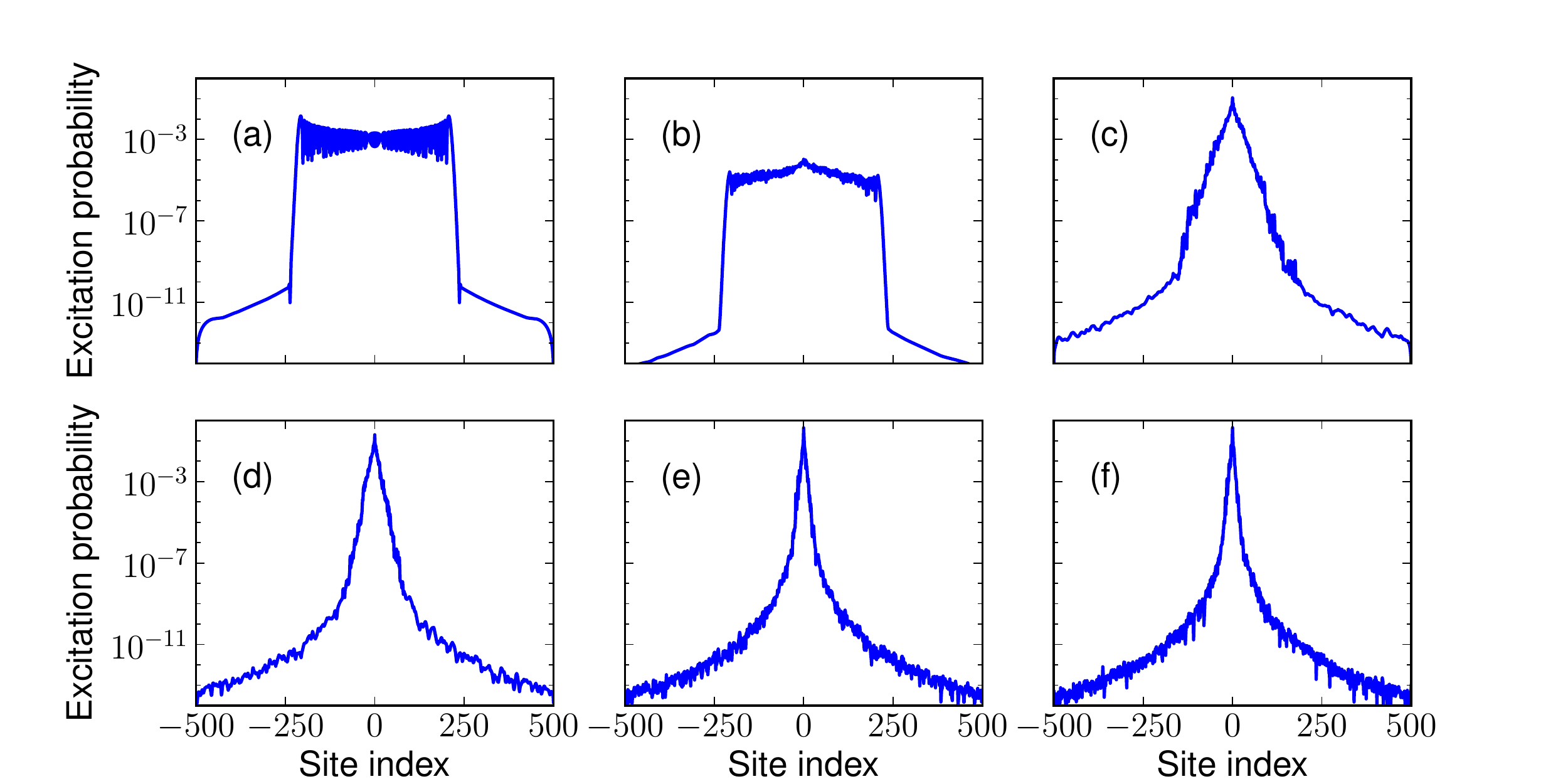}
\end{center}
\caption{
Averaged rotational excitation probability distributions formed at $t=2$ s by a single rotational excitation placed at $t=0$ in site $n=0$ of a 1D lattice of KRb molecules with 
randomly distributed empty sites. The concentration of vacancies is zero (a), 1 \% (b),  10 \% (c), 20\% (d), 50\% (e) and 70\% (f). The results are averaged over $100$ realizations of disorder.
}
\end{figure}

\begin{figure}[ht]
\label{figure2}
\begin{center}
\includegraphics[scale=0.7]{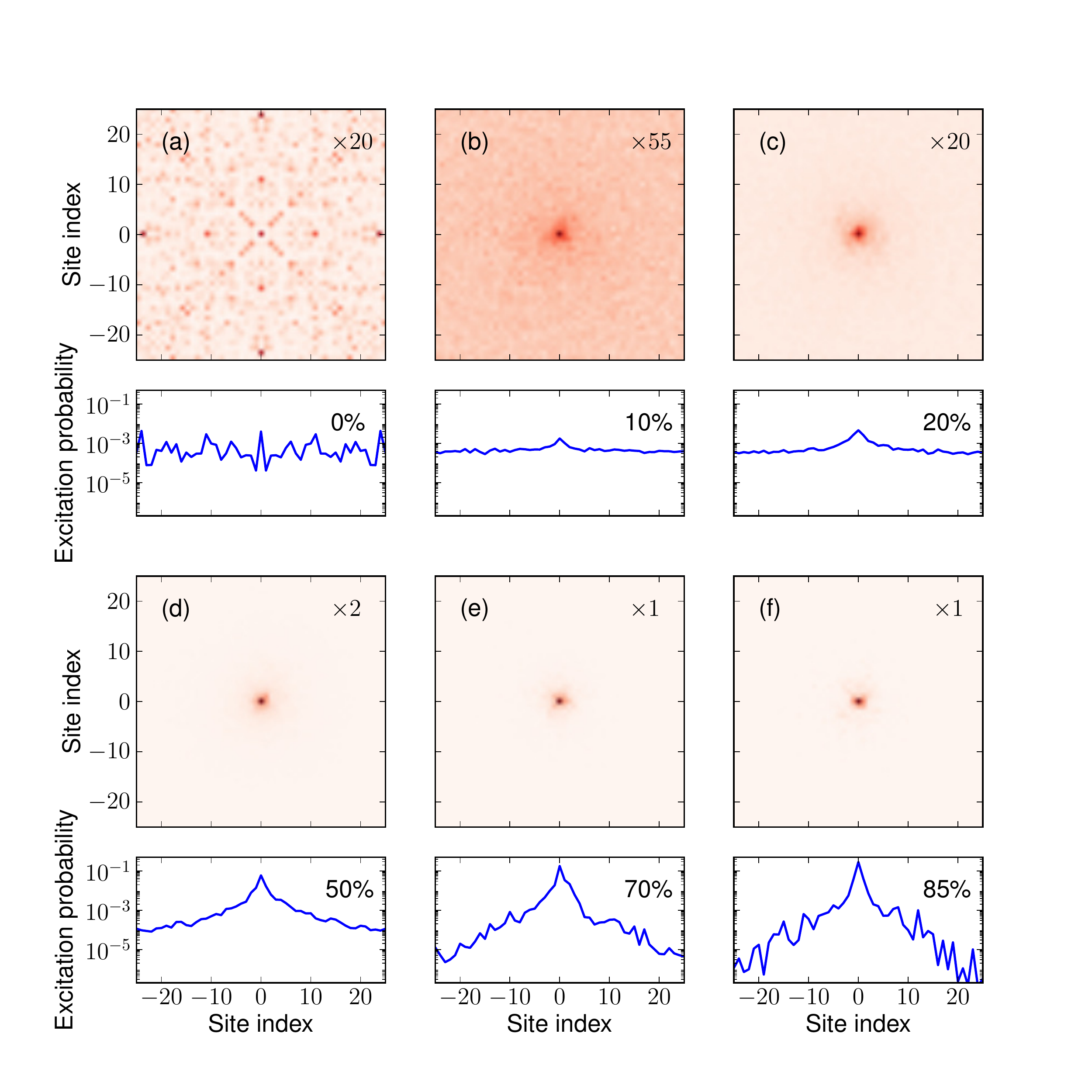}
\end{center}
\caption{
Averaged rotational excitation probability distributions formed at $t=5$ s by a single rotational excitation placed at $t=0$ in site $n_x=0, n_y = 0$ of a 2D lattice of KRb with a total of $51\times51$ sites containing a 
random distribution of empty sites. The concentration of vacancies is zero (a), 10 \% (b),  20 \% (c), 50\% (d), 70\% (e) and 85\% (f). The results are averaged over $100$ realizations of disorder. For better visualization, the probability values in the two-dimensional plots are magnified by the indicated factor. 
}
\end{figure}

\begin{figure}[ht]
\label{figure3}
\begin{center}
\includegraphics[scale=0.7]{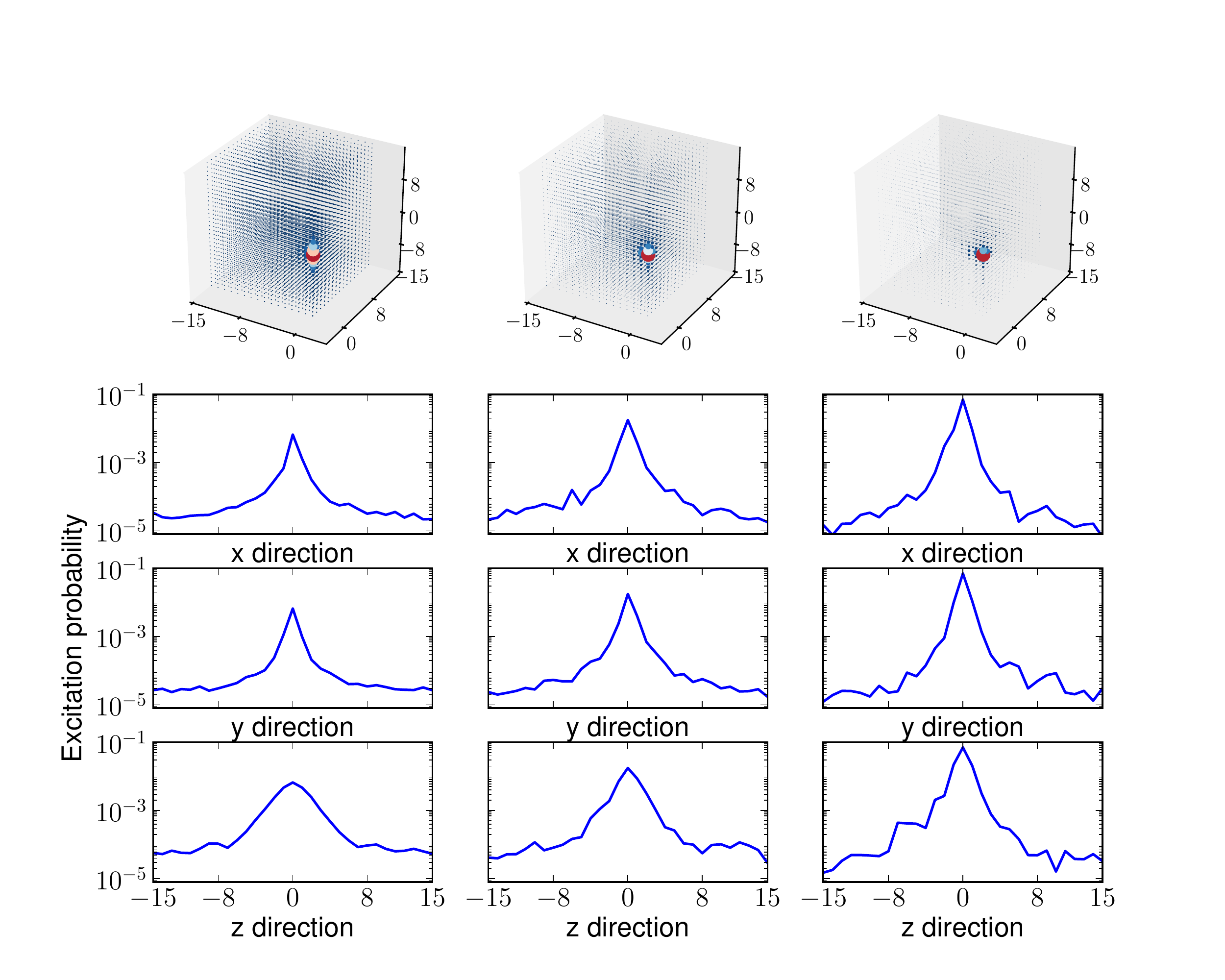}
\end{center}
\caption{
Averaged rotational excitation probability distributions formed at $t=5$ s by a single rotational excitation placed at $t=0$ in site $n_x=0, n_y = 0, n_z=0$ of a 3D lattice of KRb molecules with a total of $31\times51\times31$ sites containing a random distribution of empty sites. The concentration of vacancies is  50\% (left panels), 70\% (middle panels) and 85\% (right panels). The results are averaged over $300$ realizations of disorder. The upper panels show the part of the distributions for
$x\leq0$ and $y\geq0$.
For better visualization, the probability values in the three-dimensional plots are magnified by a factor of 27 (left), 10 (middle), and 2.5 (right). 
}
\end{figure}

\begin{figure}[ht]
\label{figure4}
\begin{center}
\includegraphics[scale=0.4]{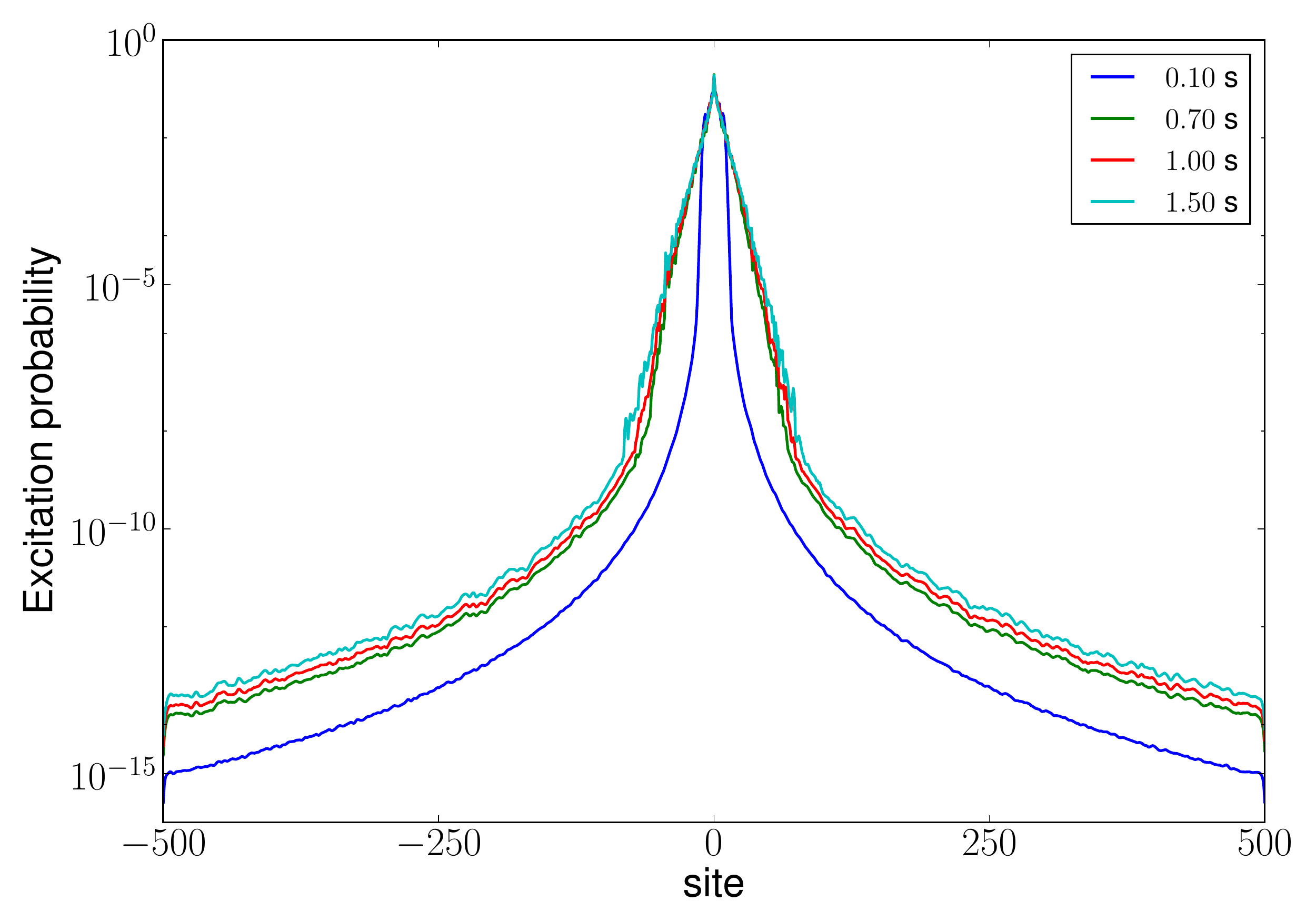}
\includegraphics[scale=0.5]{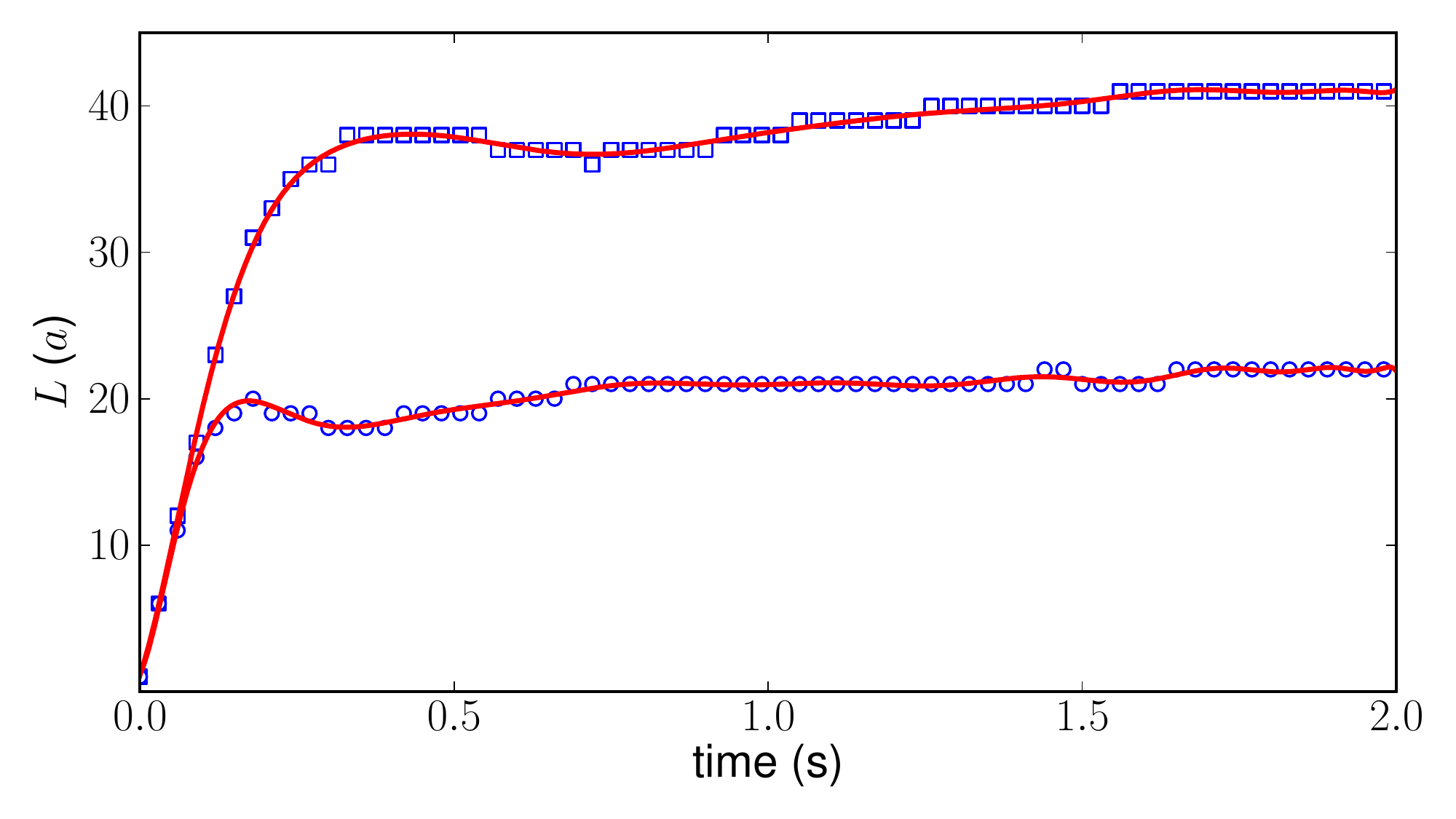}
\end{center}
\caption{
Upper panel: Time dependence of the rotational excitation probability distributions formed from one rotational excitation placed at $t=0$ in the lattice site $n=0$ of a 1D lattice of KRb molecules with 20 \% of vacancies.  Lower panel: Time dependence of the distribution width $L$ (in units of the lattice constant $a$) defined as the range of the lattice containing 90 \% of the rotational excitation probability: squares -- vacancy concentration 10 \%; circles -- vacancy concentration 20 \%. 
The results at each time are averaged over 1000 random realizations of the vacancy distributions. 
}
\end{figure}

\begin{figure}[ht]
\label{figure5}
\begin{center}
\includegraphics[scale=0.7]{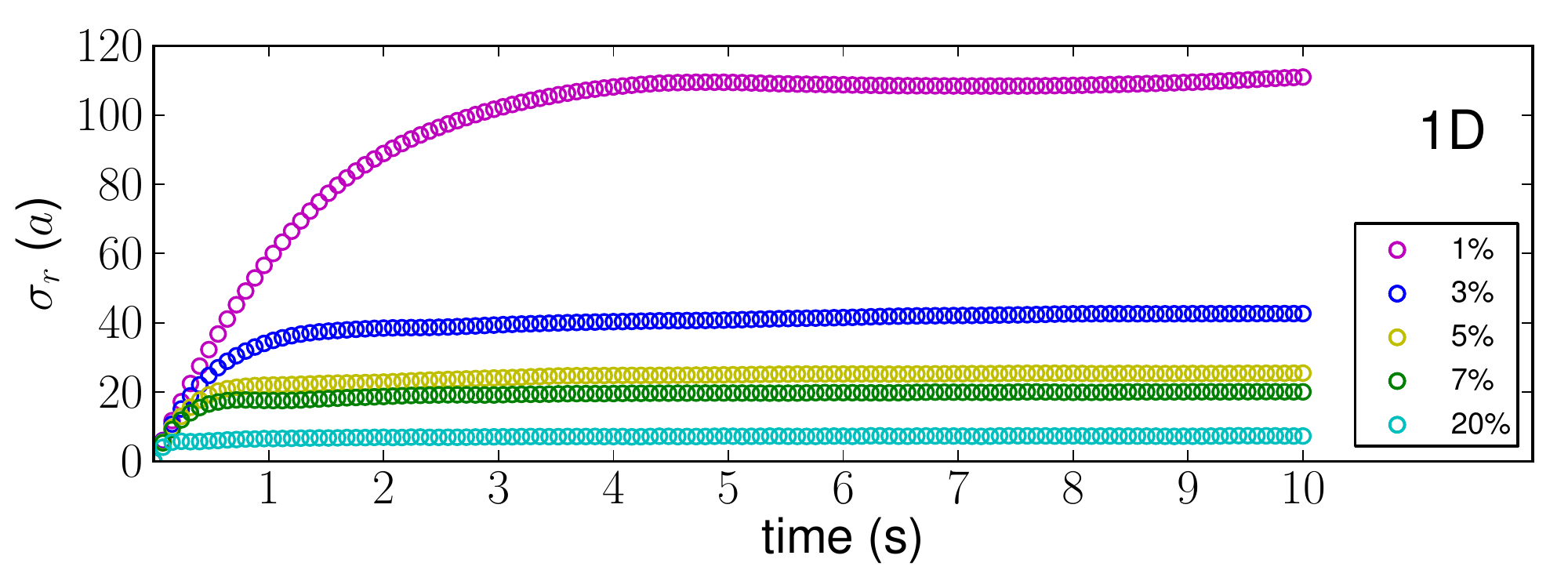}
\includegraphics[scale=0.7]{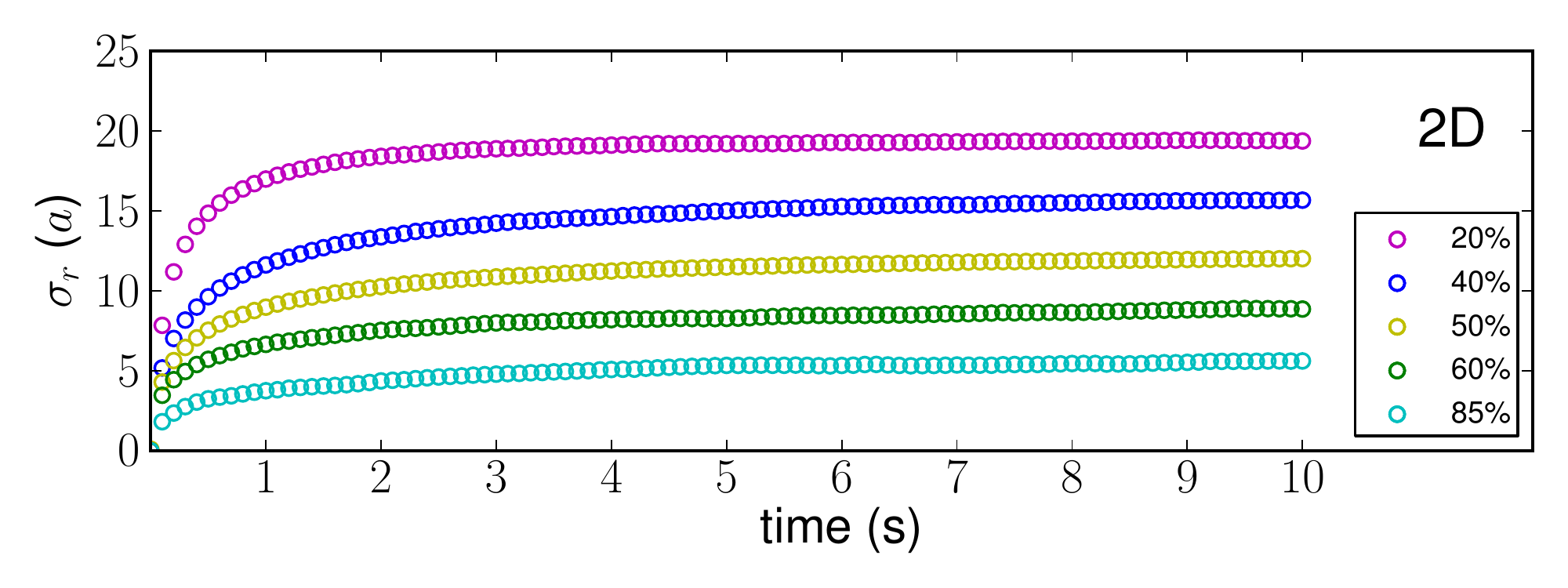}
\includegraphics[scale=0.7]{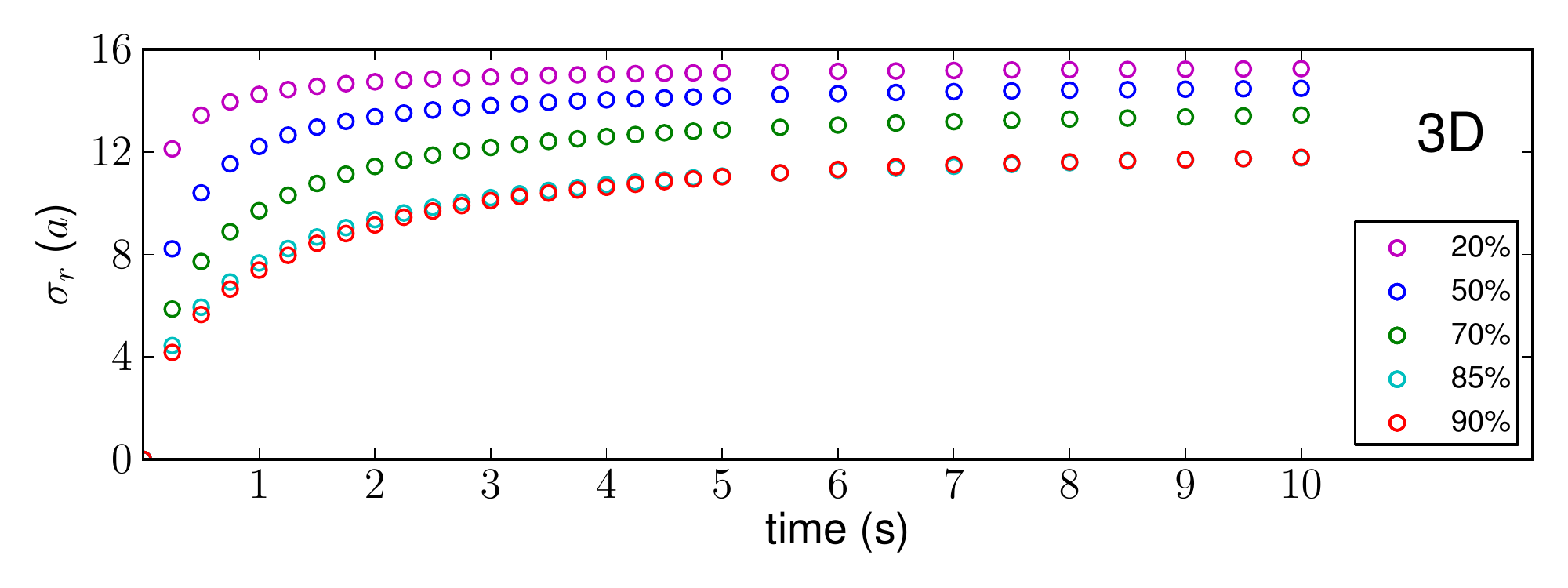}
\end{center}
\caption{
Time dependence of the standard deviations $\sigma_r$ of the rotational excitation probability distributions formed from one rotational excitation placed at $t=0$ in the centre of a lattice partially populated with molecules. Upper panel: 1D lattice with 1001 sites; middle panel: 2D square lattice with 51 $\times$ 51 lattice sites; lower panel: 3D cubic lattice with 31 $\times$ 31 $\times$ 31 lattice sites. 
The results at each time are averaged over $> 100$ random realizations of disorder.
The different sets of data correspond to different concentrations of empty lattice sites. }
\end{figure}

\begin{figure}[ht]
\label{figure6}
\begin{center}
\includegraphics[scale=0.7]{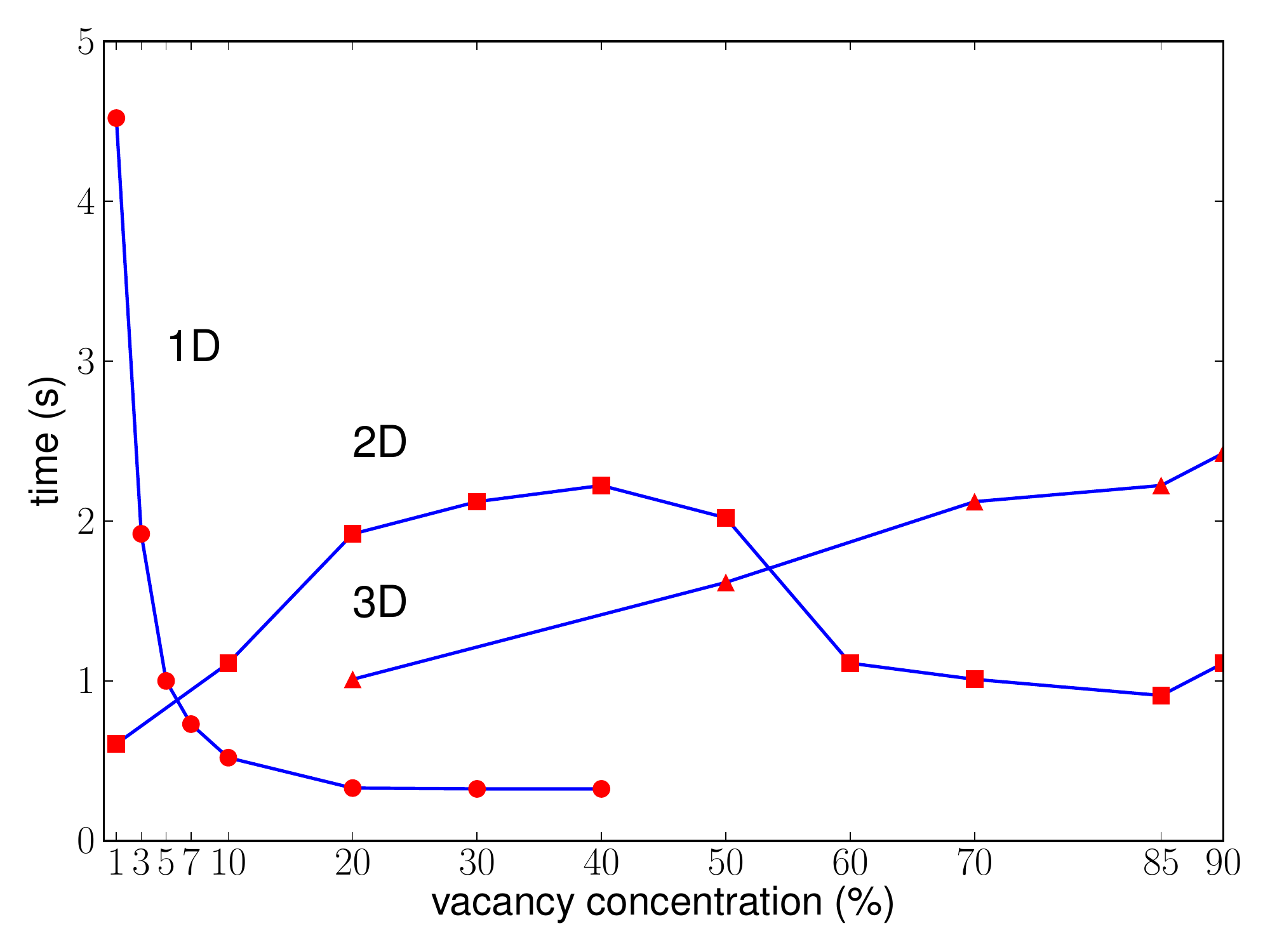}
\end{center}
\caption{
Minimum time $\tau$ required for a single rotational excitation to form a time-independent averaged distribution as a function of vacancy concentration
 in 1D (circles),  2D (squares) and 3D (triangles) disordered lattices as a function of vacancy concentration. The results at each vacancy concentration are averaged over $>100$ realizations of disorder. 
  }
\end{figure}

\begin{figure}[ht]
\label{figure7}
\begin{center}
\includegraphics[scale=0.45]{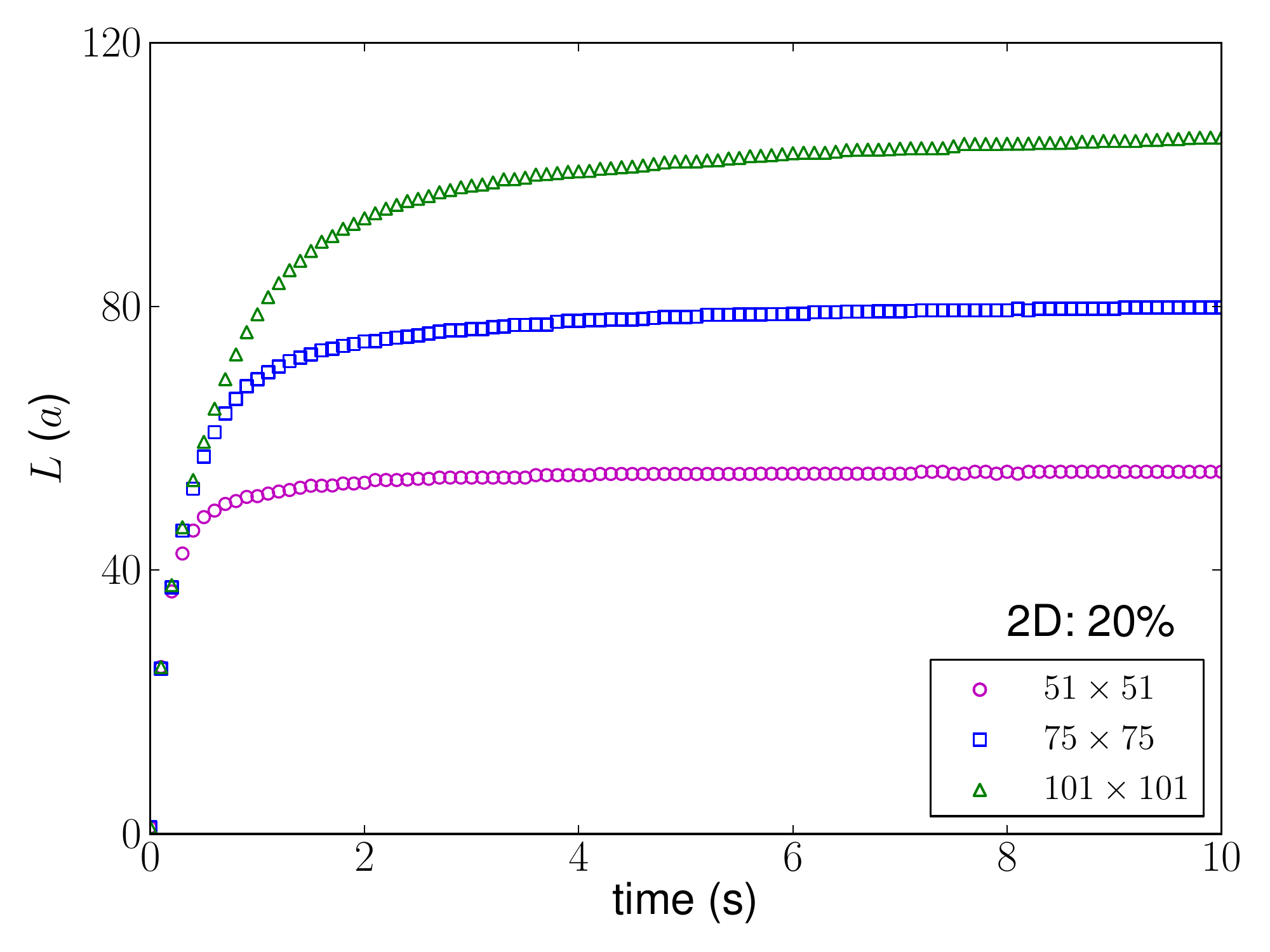}
\includegraphics[scale=0.45]{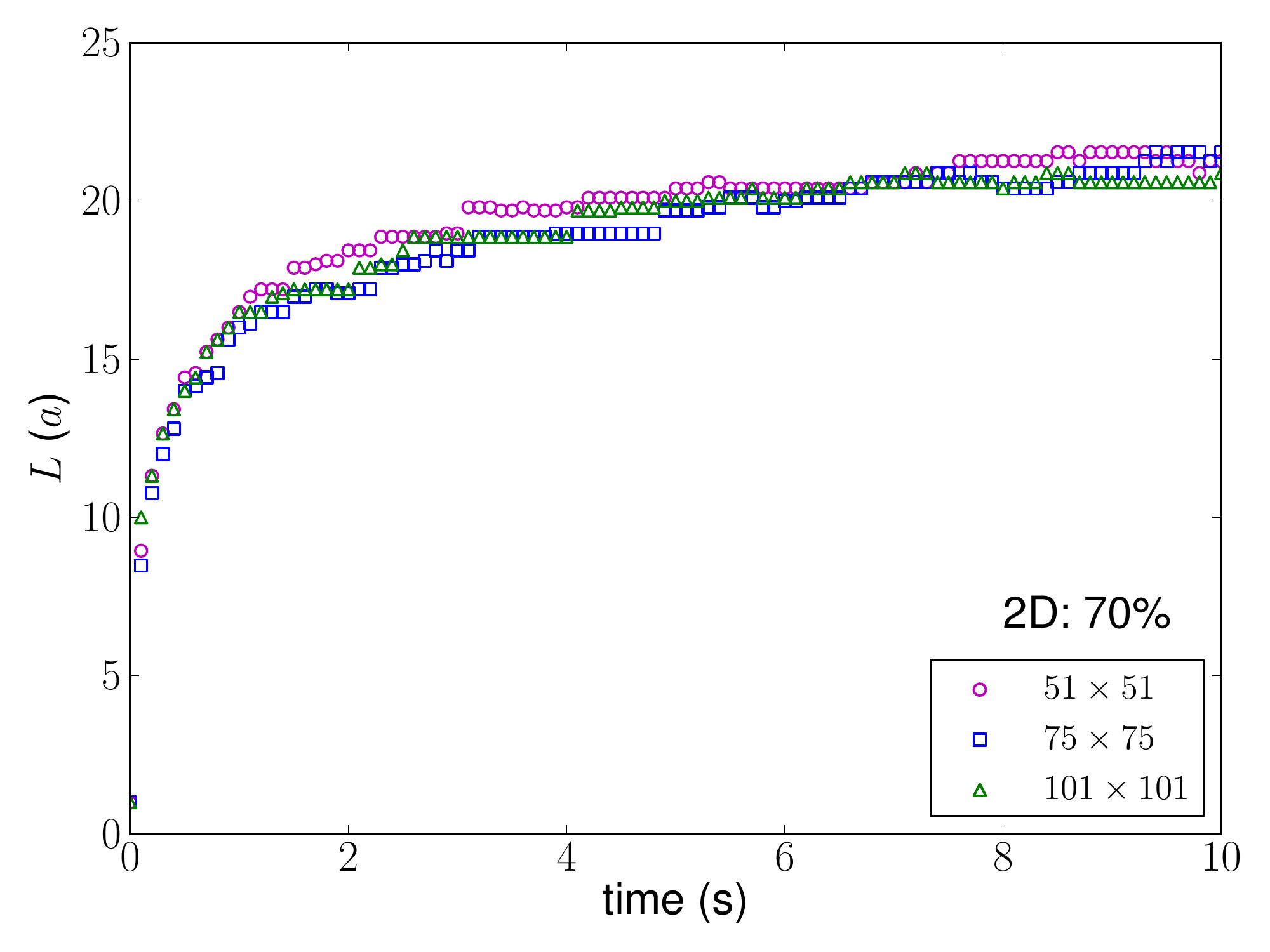}
\includegraphics[scale=0.45]{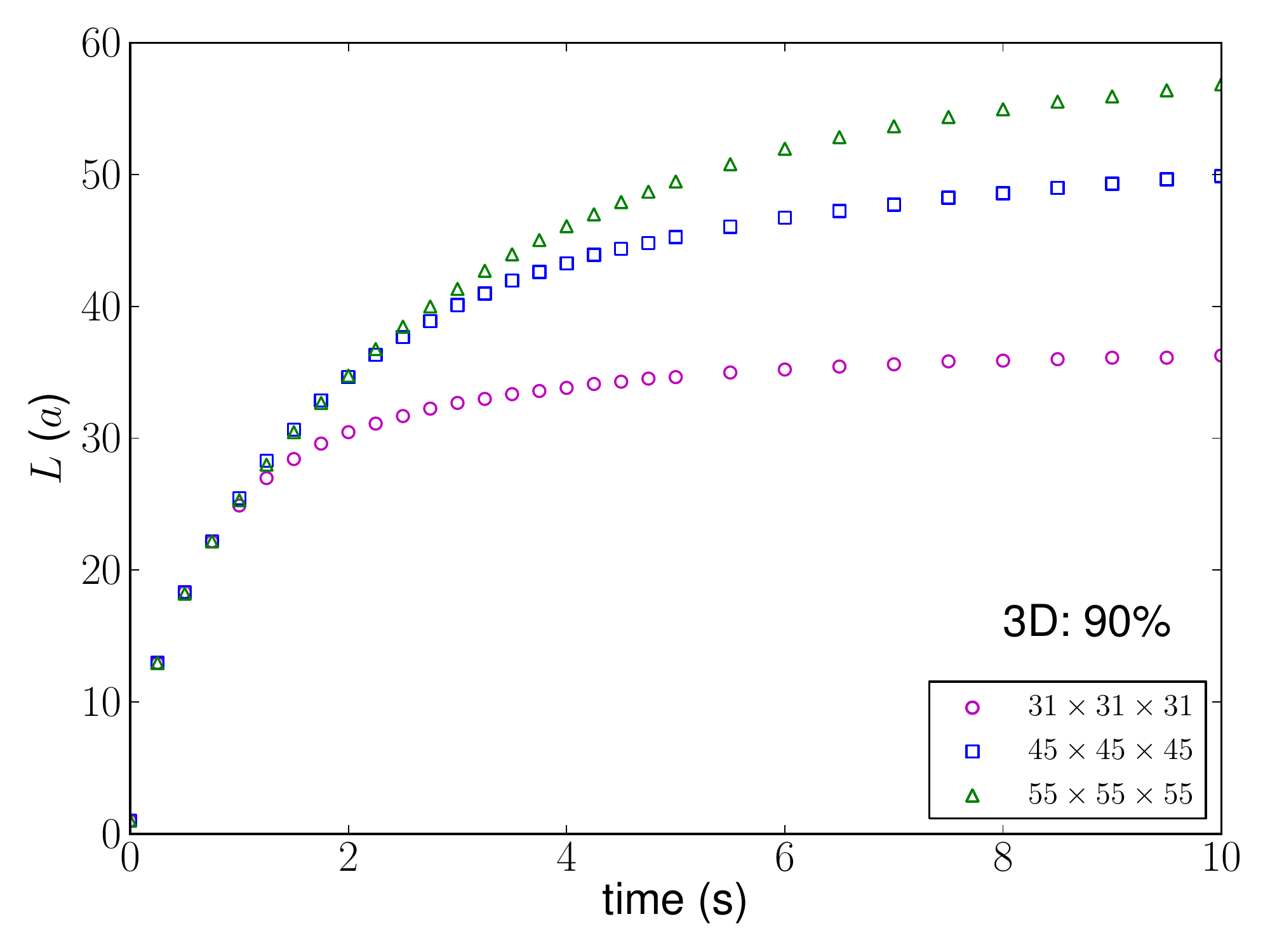}
\end{center}
\caption{
Time dependence of the width $L$ (in units of the lattice constant $a$)  of the averaged rotational excitation distributions formed by one rotational excitation placed at $t=0$ in the centre of 2D and 3D  lattices with different size. The upper and the lower panels show that the excitations diffuse to the edges of the lattice, while the middle plane illustrates that the excitation is localized within the lattice and is not affected by the lattice boundaries. 
}
\end{figure}

\begin{figure}[ht]
\label{figure8}
\begin{center}
\includegraphics[scale=0.40]{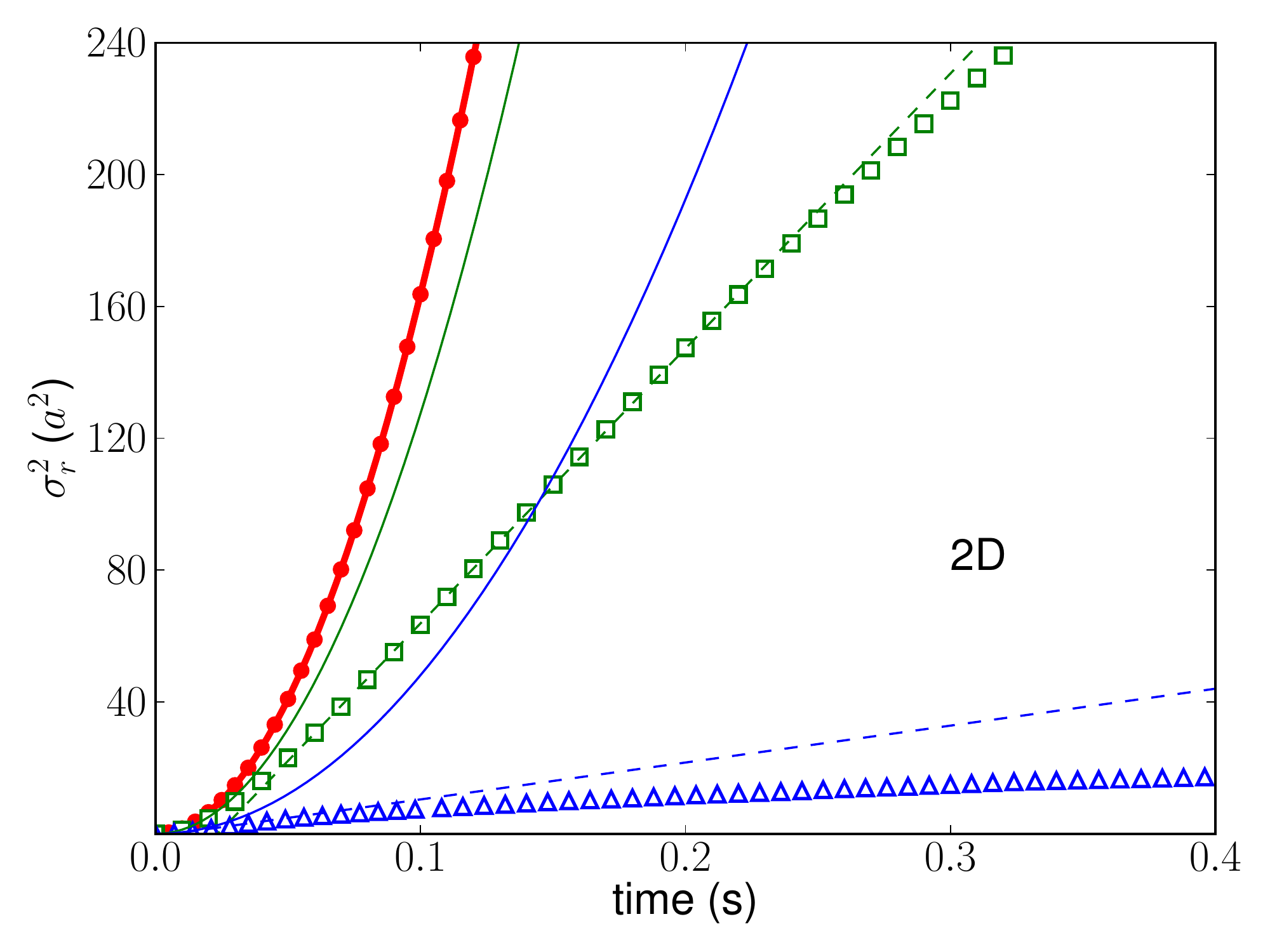} 
\includegraphics[scale=0.40]{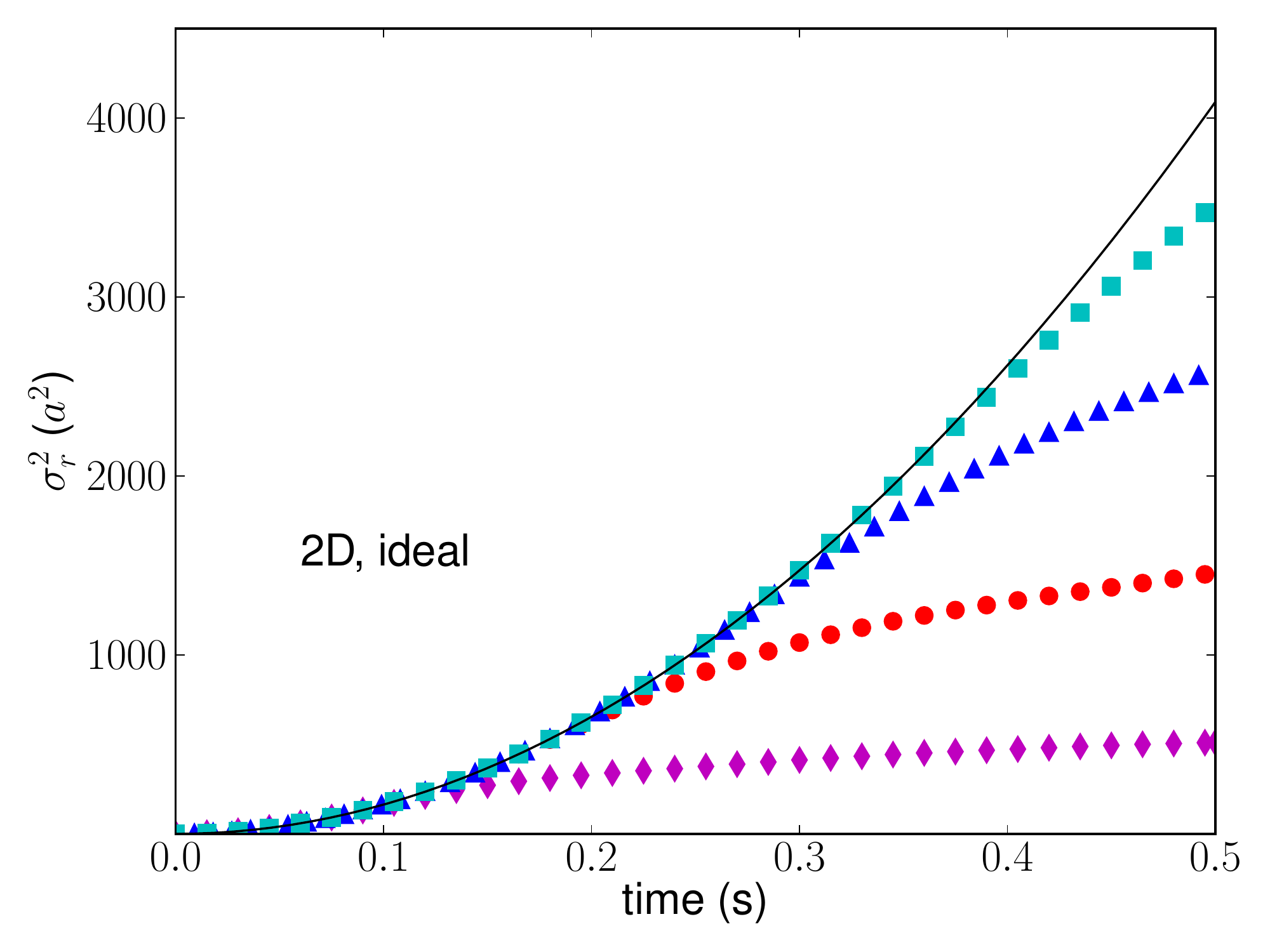} \\
\vskip-160.pt \hspace{7.7cm}
\includegraphics[scale=0.37]{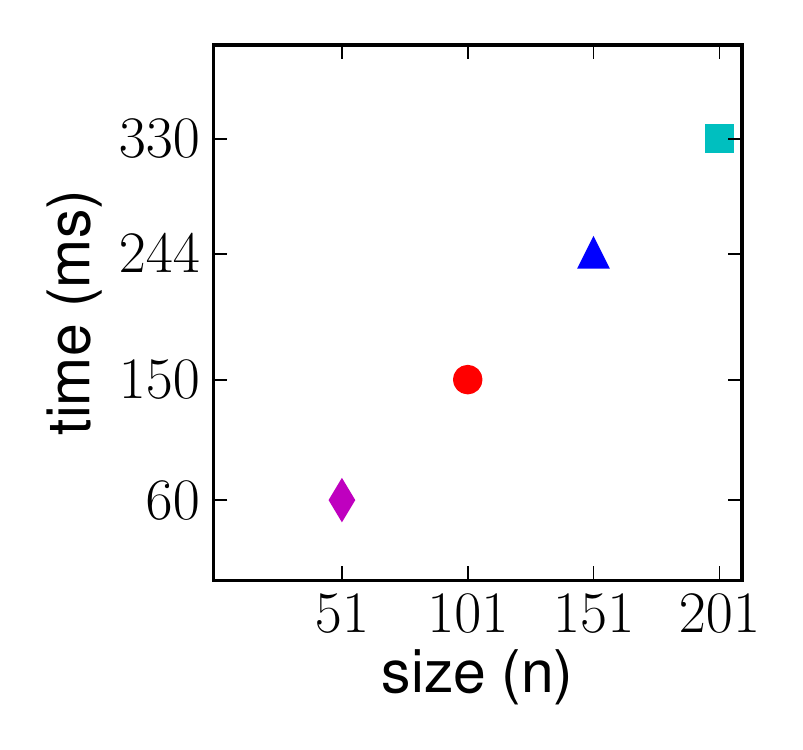} 
\vskip100.pt
\includegraphics[scale=0.45]{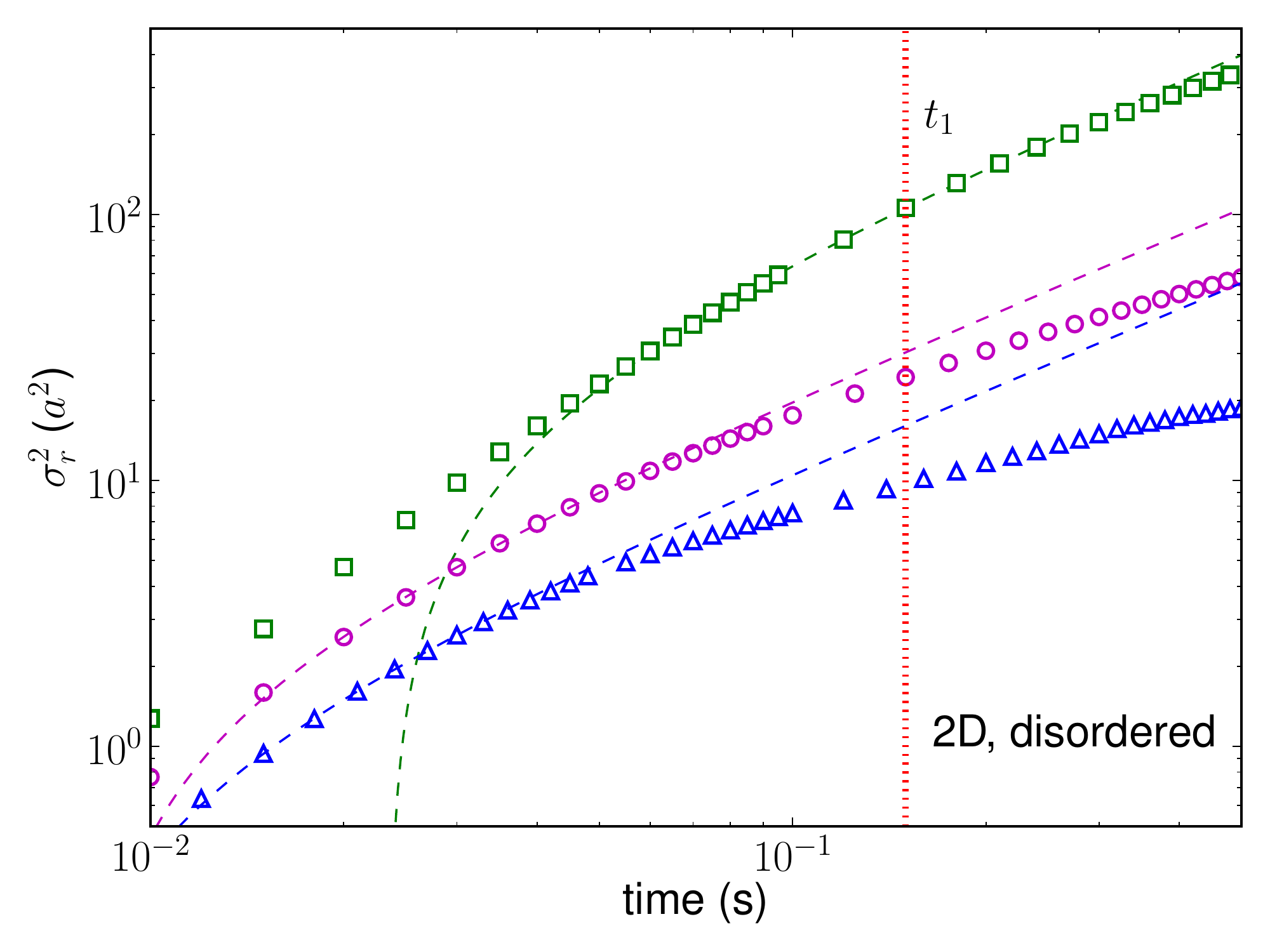}
\end{center}
\caption{Time dependence of $\sigma^2_r = \langle r^2 \rangle - \langle r \rangle^2$ for a rotational excitation initially placed in the middle of a 2D lattice.
Upper left panel: circles -- ideal lattice; squares -- lattice with 20 \% of vacancies; triangles --  lattice with 50 \% of vacancies.
Upper right panel: ideal lattice with size 51 $\times$ 51 (diamonds); 101 $\times$ 101 (circles); 151 $\times$ 151 (triangles); 201 $\times$ 201 (squares).  
Inset: Lattice size dependence of time $t_1$ marking the deviation of the wave packet dynamics from that in an ideal infinite lattice due to the boundary effects. 
Lower panel: 2D lattice with 20 \% of vacancies (squares); 50 \% of vacancies (circles); 70 \% of vacancies (triangles).
The symbols represent the numerical calculations; the full curves are the analytical fits  $\sigma^2_r = D t^2$; the dashed curves are the linear fits  $\sigma^2_r = b t + c$. 
The size of the lattice for the results in the upper left panel and the lower panel is $101 \times 101$. 
}
\end{figure}

\begin{figure}[ht]
\label{figure9}
\begin{center}
\includegraphics[scale=0.40]{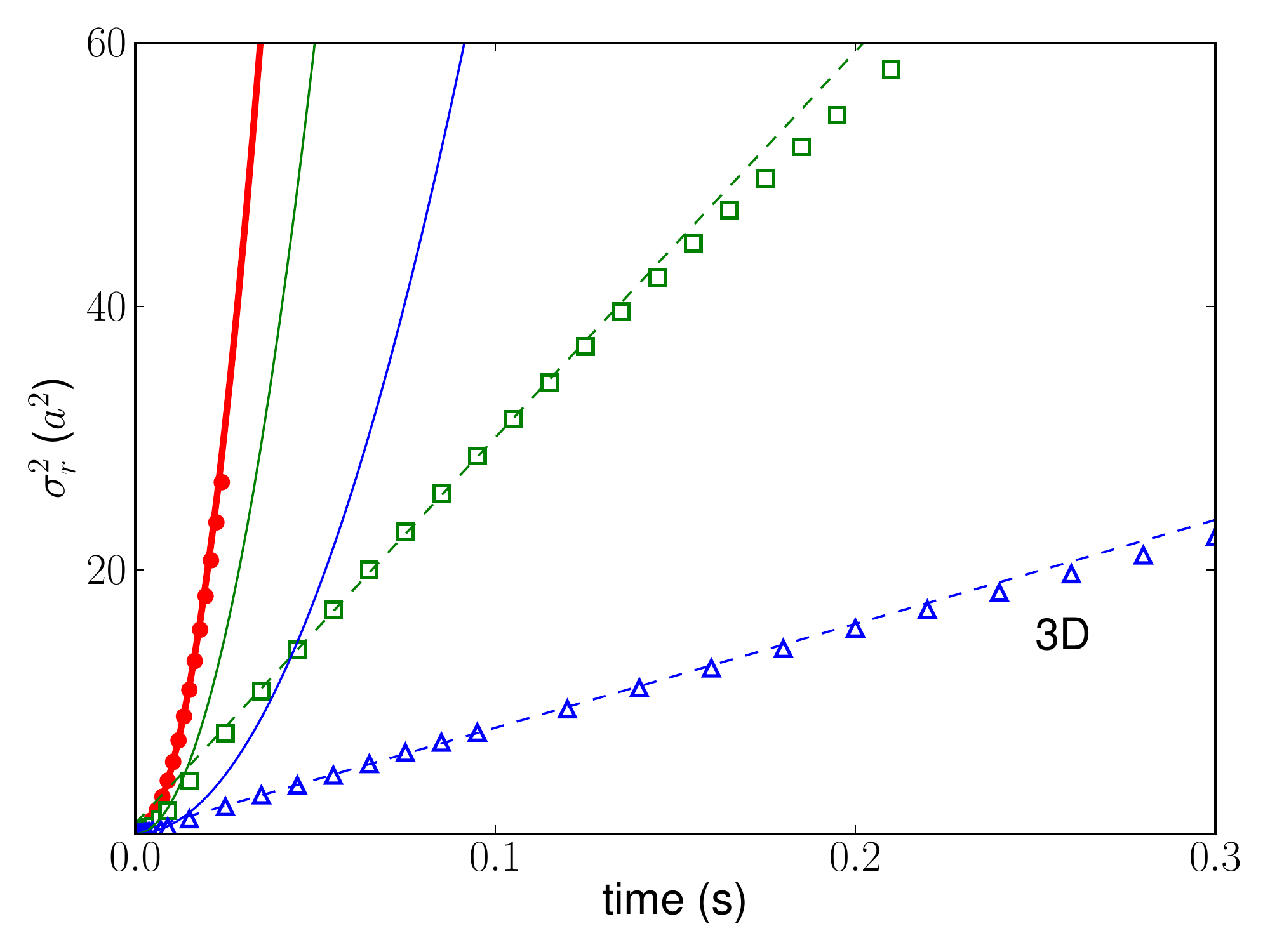} 
\includegraphics[scale=0.40]{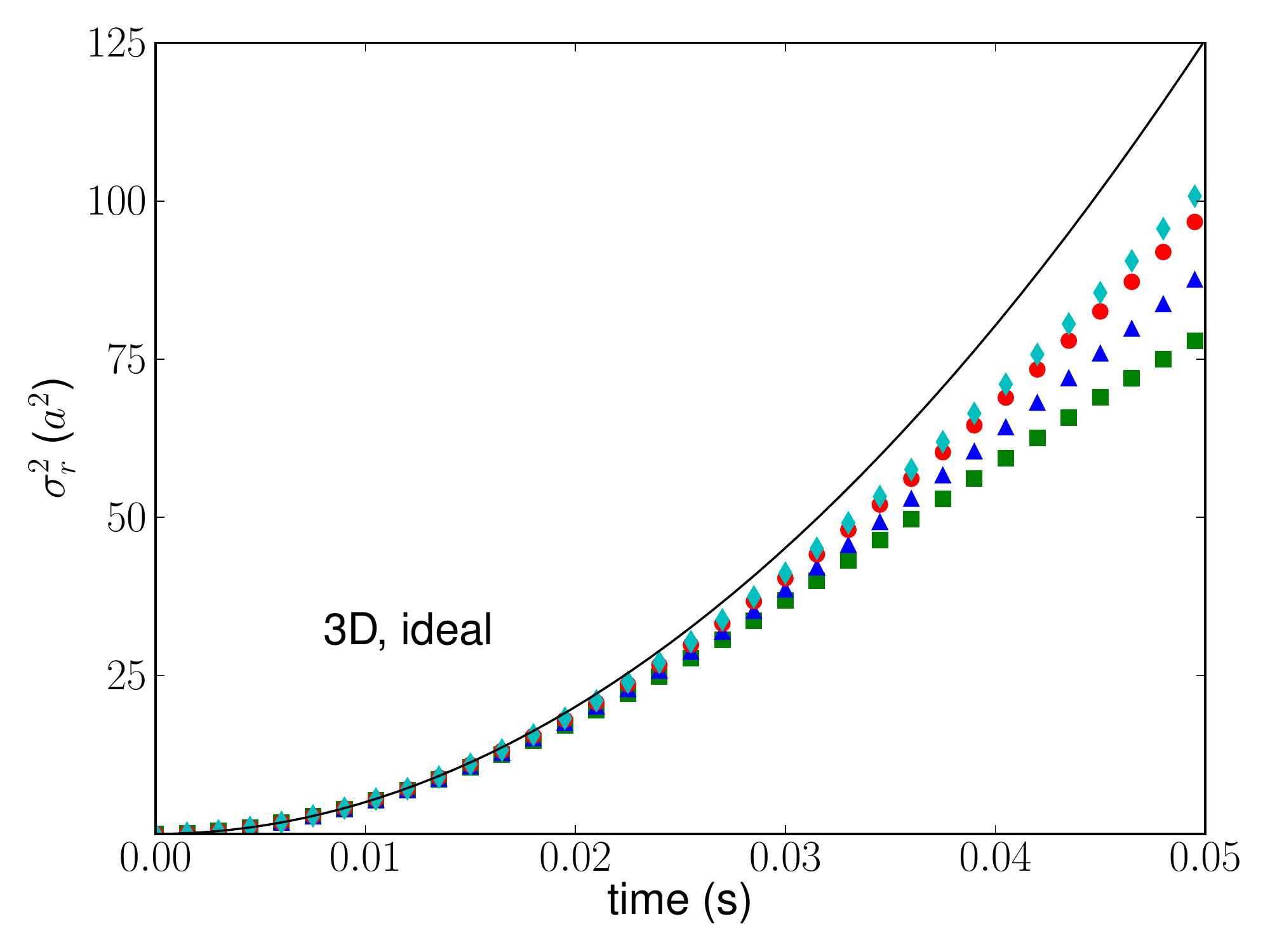} \\
\vskip-160.pt \hspace{7.7cm}
\includegraphics[scale=0.37]{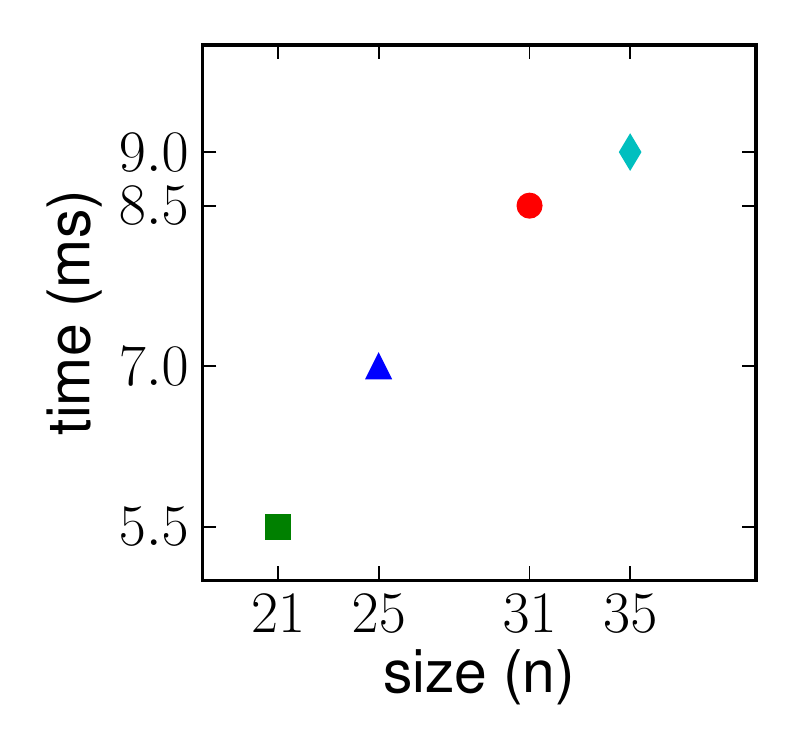} 
\vskip100.pt
\includegraphics[scale=0.45]{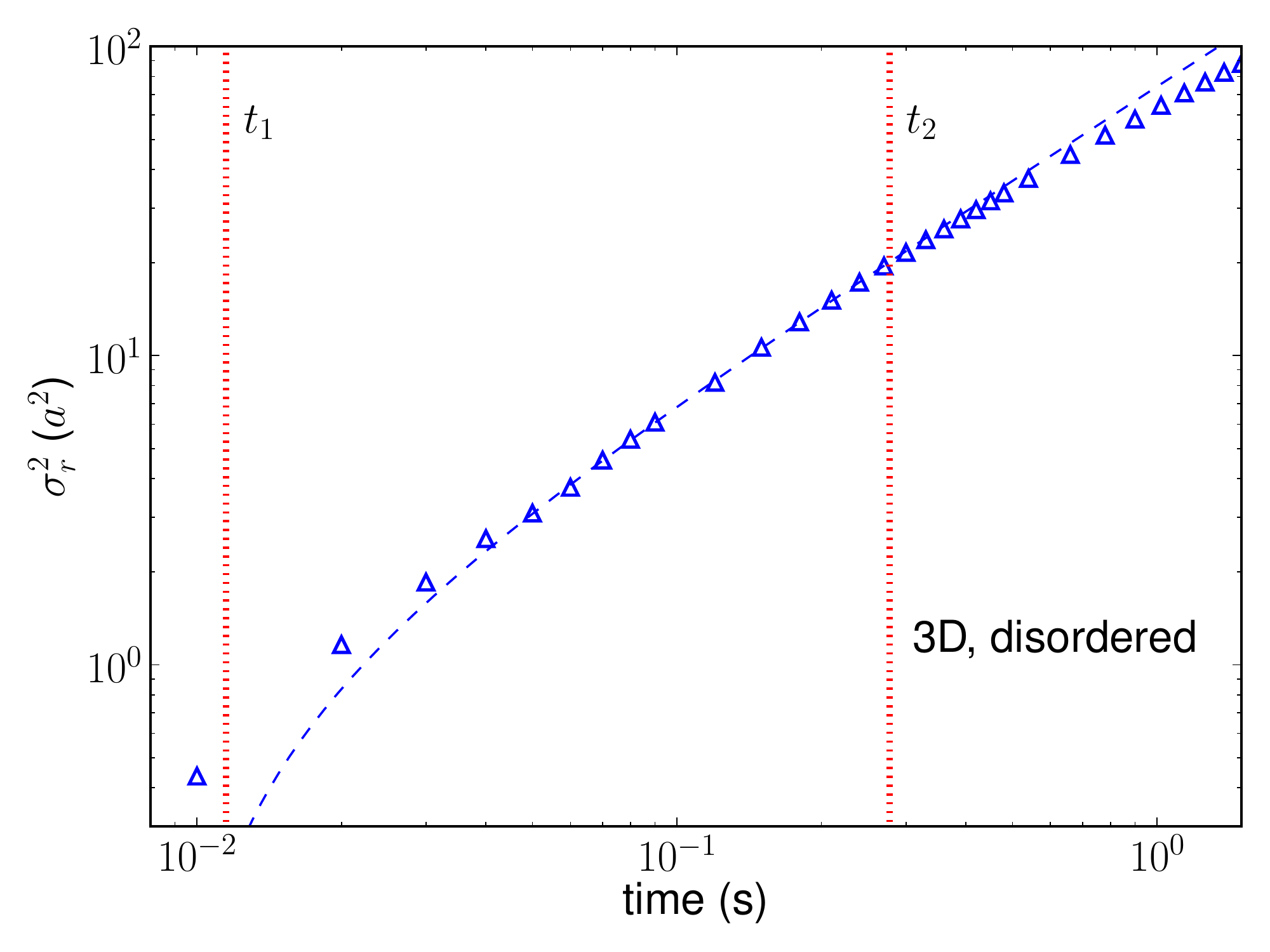}
\end{center}
\caption{Time dependence of $\sigma^2_r = \langle r^2 \rangle - \langle r \rangle^2$ for a rotational excitation initially placed in the middle of a 3D lattice.
Upper left panel: 3D ideal lattice (circles), disordered lattice with 50 \% of vacancies (squares) and disordered lattice with 85 \% of vacancies with size $31 \times 31 \times 31$ sites. 
Upper right panel: 3D ideal lattice with size 21 $\times$ 21 $\times$ 21 (squares); 25 $\times$ 25 $\times$ 25 (triangles); 31 $\times$ 31$\times$ 31 (circles); 35 $\times$ 35 $\times$ 35 (diamonds).
Inset: Lattice size dependence of time $t_1$ marking the deviation of the wave packet dynamics from that in an ideal infinitely lattice due to the boundary effects.   
Lower panel: 3D disordered lattice of size $55 \times 55 \times 55$ with 90 \% of vacancies. 
The symbols represent the numerical calculations; the full curves are the analytical fits  $\sigma^2_r = D t^2$; the dashed curves are the linear fits  $\sigma^2_r = b t + c$. 
}
\end{figure}


\begin{figure}[ht]
\label{figure10}
\begin{center}
\includegraphics[scale=0.5]{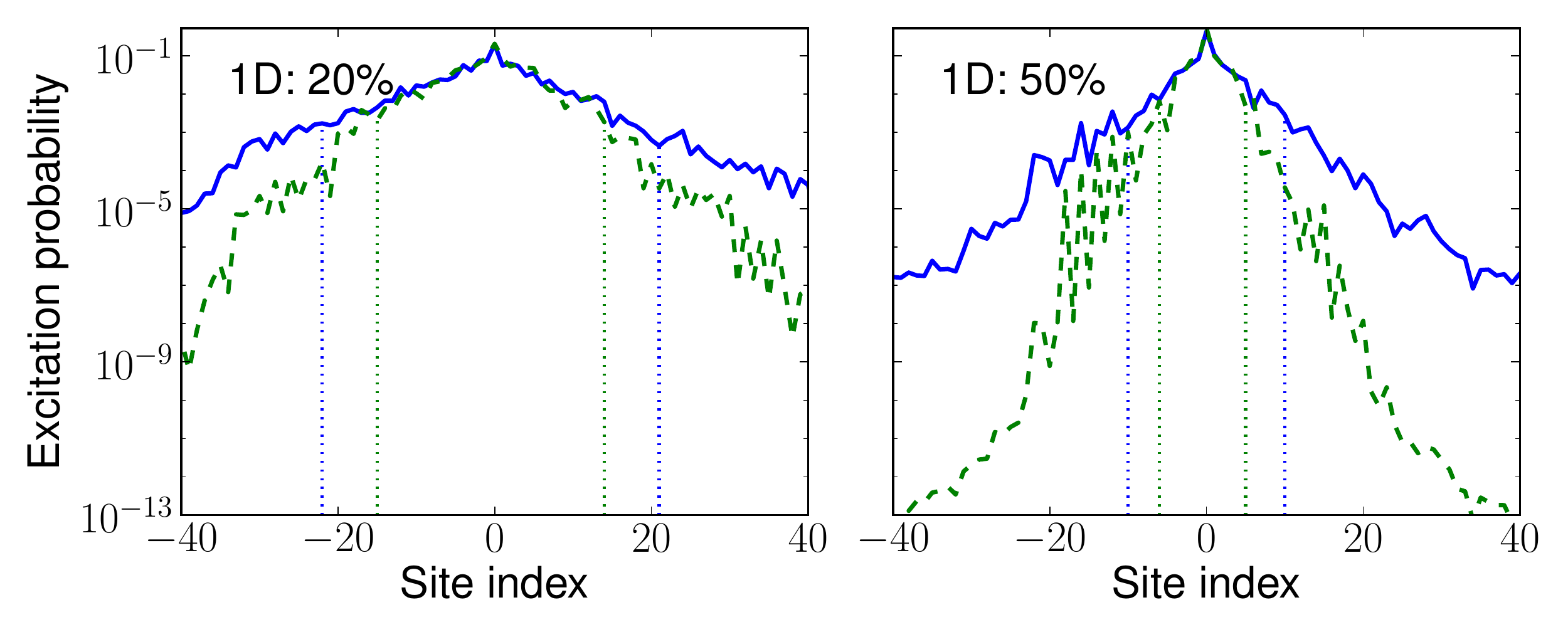}
\includegraphics[scale=0.5]{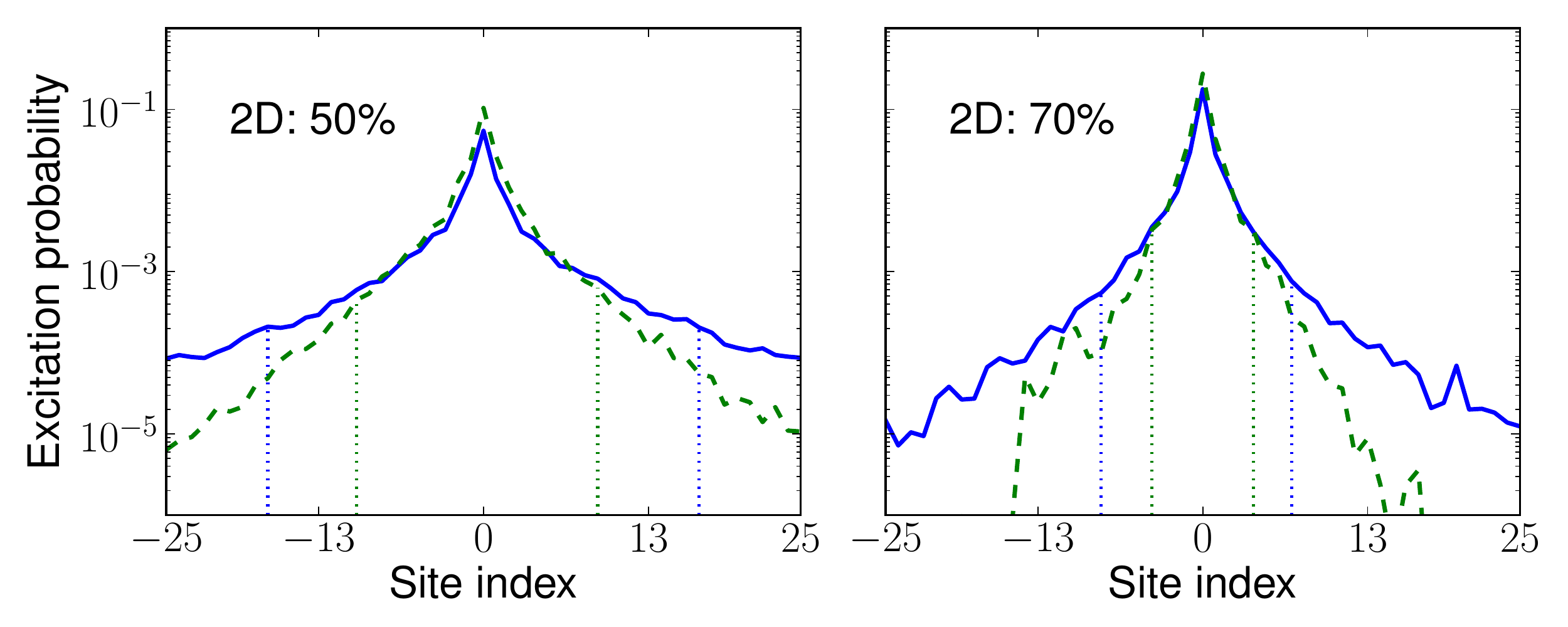}
\includegraphics[scale=0.5]{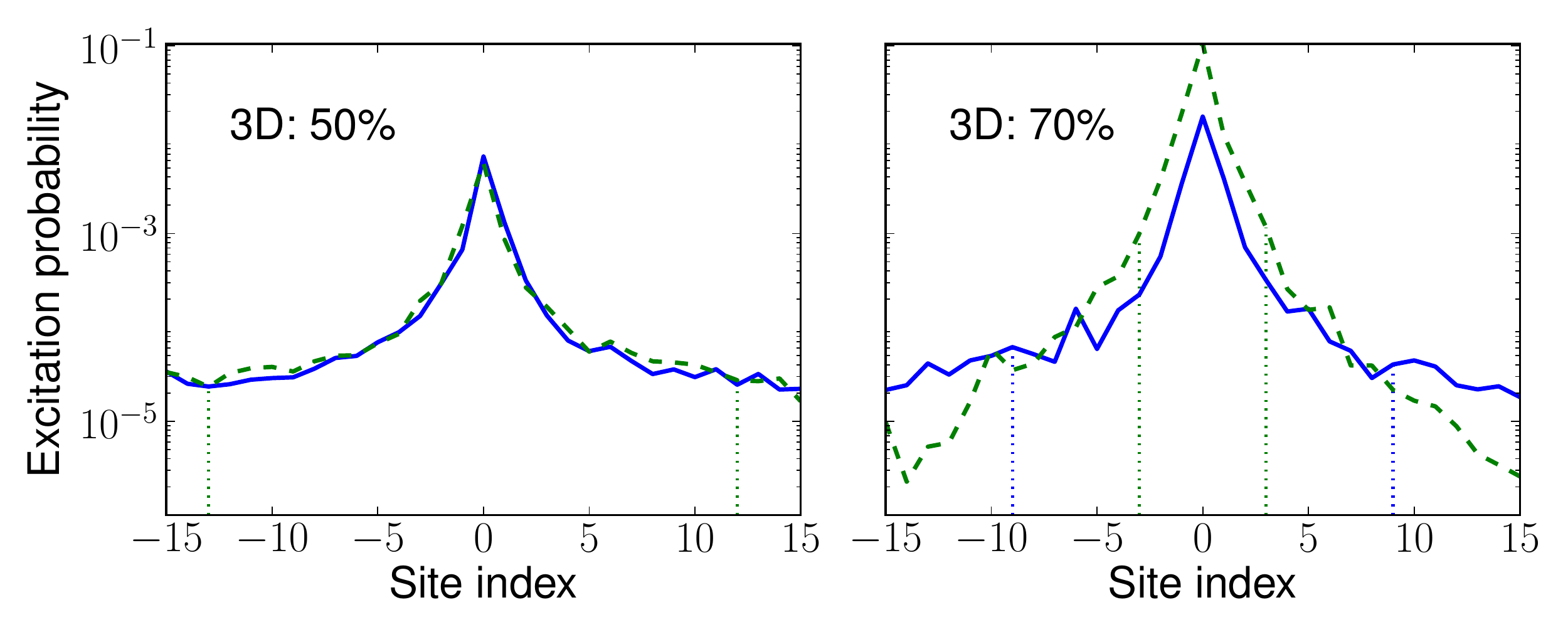}
\end{center}
\caption{
Averaged rotational excitation probability distributions formed at $t=2$ s (for 1D lattices) and $t=5$ s (for 2D and 3D lattices) by a single rotational excitation placed at $t=0$ in the center of the lattices. 
The solid curves are the results of the calculation with the tunnelling amplitudes $t_{\bm n, \bm n'}$ as defined in Eq. (\ref{t-elements}). The dashed curves are the result of the calculation with the tunnelling amplitudes  $\tilde t_{\bm n, \bm n'} \propto 1/|\bm n - \bm n'|^5$. 
The middle and lower panels show the cross sections of the 2D and 3D distributions along the $x$-axis. 
The vertical dotted lines show the distribution widths containing 99 \% of the rotational excitation. The dimensions of the lattices are 1001 sites for 1D; $51 \times 51$ sites for 2D; and $31 \times 31 \times 31$ sites for 3D.  
}
\end{figure}

\begin{figure}[ht]
\label{figure11}
\begin{center}
\includegraphics[scale=0.7]{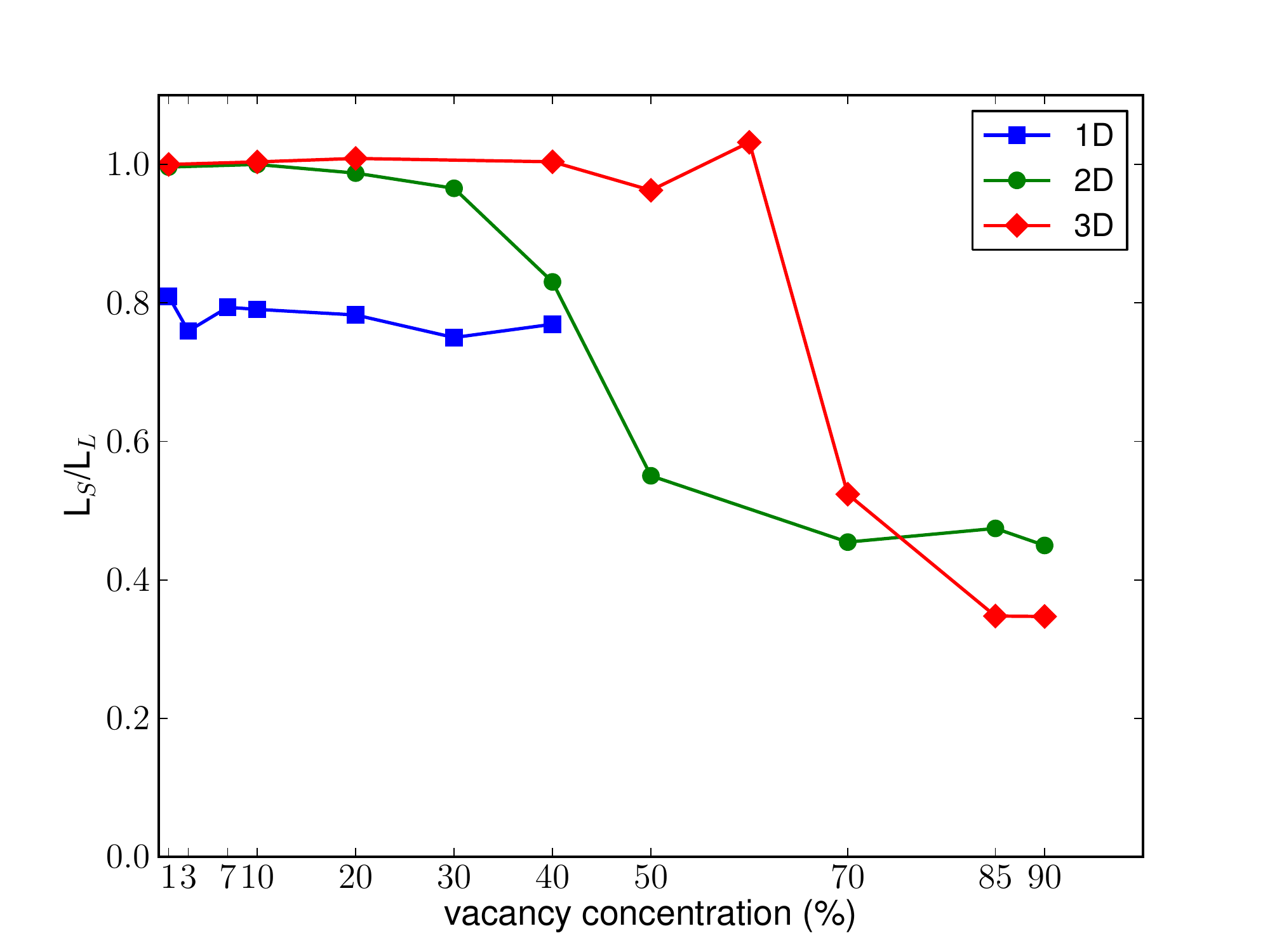}
\end{center}
\caption{
The ratio of the widths of the rotational excitation distributions $L_S/L_L$. The distribution widths $L_L$ are computed 
with the long-range amplitudes $t_{\bm n, \bm n'}$ as defined in Eq. (\ref{t-elements}). 
The distribution widths $L_S$ are computed 
with the short-range amplitudes $\tilde t_{\bm n, \bm n'} \propto 1/|\bm n - \bm n'|^5$.
The results at each vacancy concentration are averaged over $>100$ realizations of disorder. 
}
\end{figure}

\begin{figure}[ht]
\label{figure12}
\begin{center}
\includegraphics[scale=0.45]{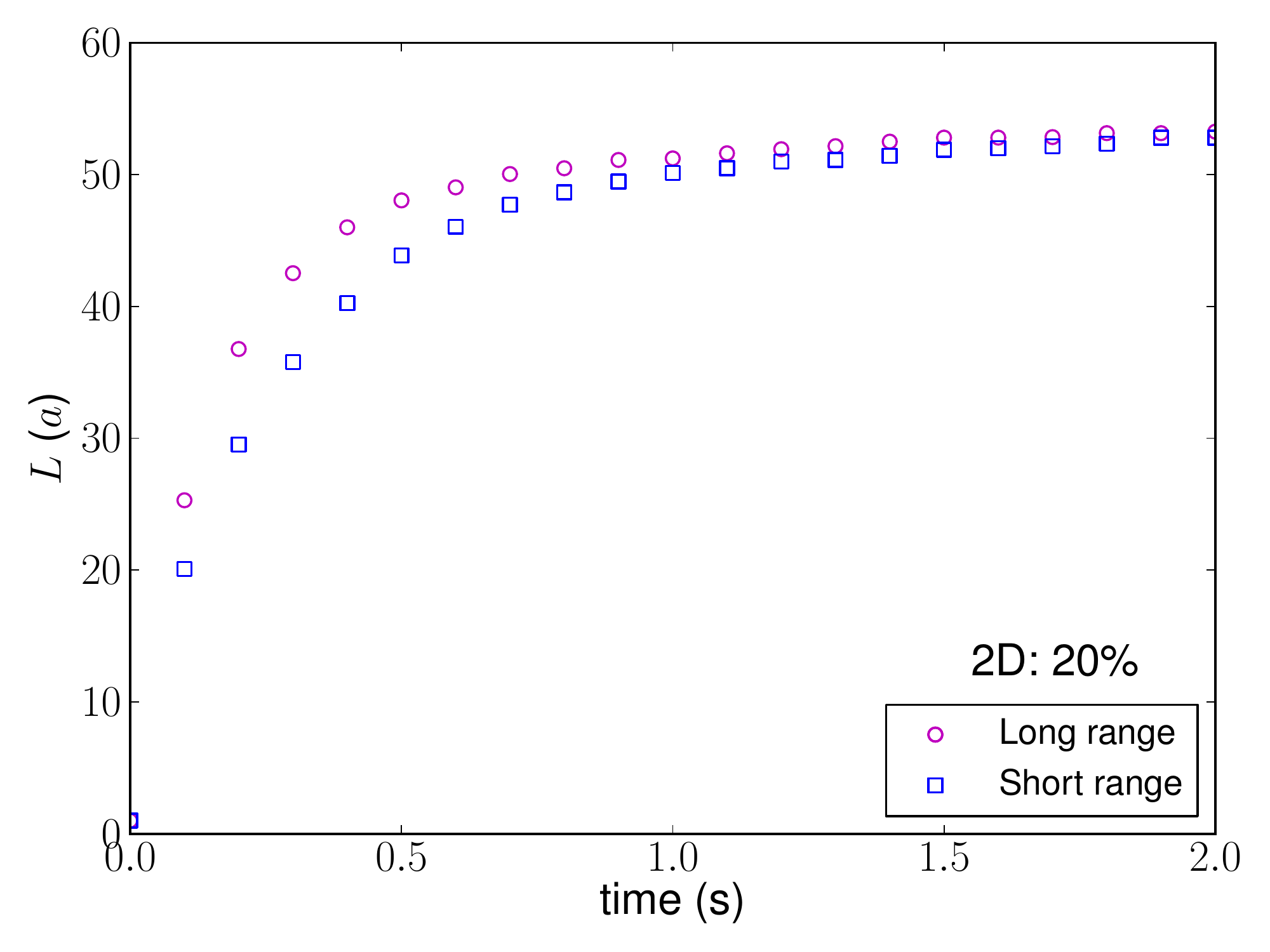}
\includegraphics[scale=0.45]{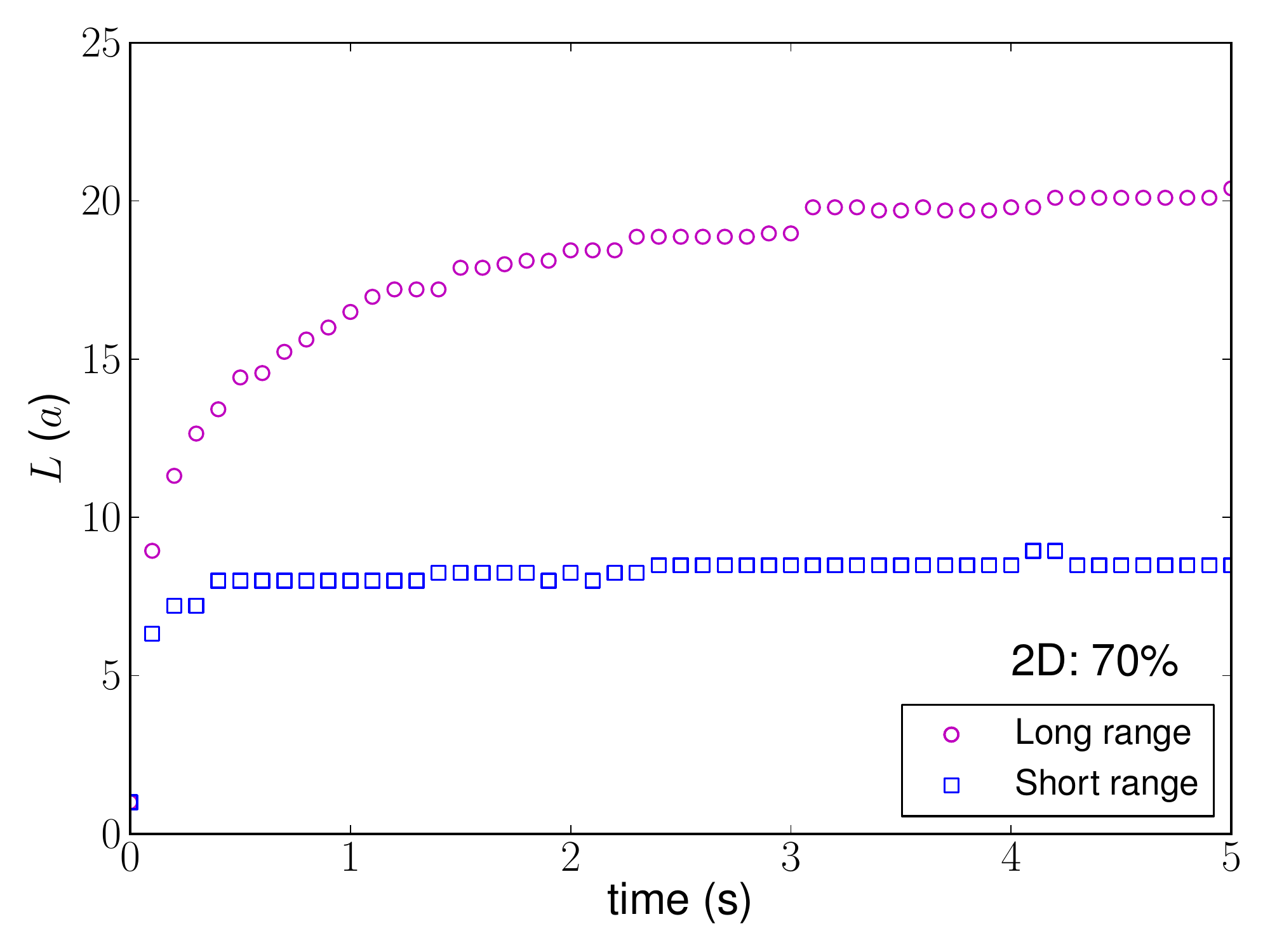}
\end{center}
\caption{
Time dependence of the width $L$ (in units of the lattice constant $a$)  of the averaged rotational excitation distributions formed by one rotational excitation placed at $t=0$ in the centre of 2D  lattices with different concentrations of empty lattice sites: circles -- the results computed 
with the long-range amplitudes $t_{\bm n, \bm n'}$ as defined in Eqs. (\ref{t-elements}) and (\ref{eq:vdd}); squares -- the results computed 
with the short-range amplitudes $\tilde t_{\bm n, \bm n'} \propto 1/|\bm n - \bm n'|^5$. 
}
\end{figure}


\clearpage
\newpage

\begin{thebibliography}{99}




\bibitem{anderson}
P. W. Anderson, {\it Phys. Rev.} {\bf 109}, 1492 (1958).

\bibitem{review-rmp}
F. Evers and A. D. Mirlin
{\it Rev. Mod. Phys.} {\bf 80}, 1355 (2008). 

\bibitem{example-3}
M. Imada, A. Fujimori, and Y. Tokura,
{\it Rev. Mod. Phys.} {\bf 70}, 1039 (1998).

\bibitem{example-4}
C. Bauer and H. Giessen,
{\it J. Opt.} {\bf 16}, 114001 (2014).

\bibitem{example-5}
B. Abeles, P. Sheng, M. D. Coutts, and Y. Arie, 
{\it Adv. Phys.} {\bf 24}, 407 (1975).

\bibitem{example-1}
P. N. Butcher, N. H. March, and M. P. Tosi (Eds.), ``{\it Physics of Low-Dimensional Semiconductor Structures}'', Springer Science, New York (1993). 


\bibitem{mapping}
K. R. A. Hazzard, M. van den Worm, M. Foss-Feig, S. R. Manmana, E. G. Dalla Torre, T. Pfau, M. Kastner, and A. M. Rey, \pra{90}{063622}{2014}.


\bibitem{microwaves}
A. A. Chabanov, M. Stoytchev, and A. Z. Genack, {\it Nature} {\bf 404}, 850 (2000). 

\bibitem{microwaves-2}
U. Kuhl, F. M. Izrailev, A. A. Krokhin and H. -J. St\"ockmann
{\it Appl. Phys. Lett.} {\bf 77}, 633 (2000).


\bibitem{photons}
D. S. Wiersma, P. Bartolini, A. Lagendijk, and R. Righini, {\it Nature} {\bf 390}, 661 (1997).

\bibitem{photons-2}
M. St\"orzer, P. Gross, C. M. Aegerter, and G. Maret, {\it Phys. Rev. Lett.}  {\bf 96}, 063904 (2006).


\bibitem{atoms-1}
D. Cl\'ement, A. F. Var\'on, J. A. Retter, L. Sanchez-Palencia, A. Aspect, and P. Bouyer,
{\it New J. Phys.} {\bf 8}, 165 (2006). 

\bibitem{atoms-2}
J. Billy, V. Josse, Z. Zuo, A. Bernard, B. Hambrecht, P. Lugan, D. Cl\'ement, L. Sanchez-Palencia, P. Bouyer, and A. Aspect, {\it Nature} {\bf 453}, 891 (2008).

\bibitem{atoms-3}
 L. Fallani, J. E. Lye, V. Guarrera, C. Fort, and M. Inguscio, {\it Phys. Rev. Lett.}  {\bf 98}, 130404 (2007). 

\bibitem{atoms-4}
G. Roati, C. D'Errico, L. Fallani, M. Fattori, C. Fort, M. Zaccanti, G. Modugno, M. Modugno, and M. Inguscio, {\it Nature} {\bf 453}, 895 (2008).



\bibitem{scaling-hypothesis}
E. Abrahams, P. W. Anderson, D. C. Licciardello, and T. V. Ramakrishnan,
{\it Phys. Rev. Lett.} {\bf 42}, 673 (1979).




\bibitem{correlations}
Y. Lahini, Y. Bromberg, D. N. Christodoulides, and Y. Silberberg, {\it Phys. Rev. Lett.} {\bf 105}, 163905 (2010).

\bibitem{dissipation-induced-localization}
S. Johri, R. Nandkishore, and R. N. Bhatt, arXiv:1405.5515. 

\bibitem{long-range-prl}
A. Rodr{\'\i}guez�, V. A. Malyshev, G. Sierra, M. A. Mart{\'\i}n-Delgado�, J. Rodr{\'\i}guez-Laguna�, and F. Dom{\'\i}nguez-Adame�, \prl{90}{027404}{2003}.


\bibitem{long-range-dispersion}
F. A. B. F. de Moura, A. V. Malyshev, M. L. Lyra, V. A. Malyshev, and F. Domínguez-Adame, 
\prb{71}{174203}{2005}.


\bibitem{long-range-effects}
N. Y. Yao, C. R. Laumann, S. Gopalakrishnan, M. Knap, M. M\"ueller, E. A. Demler, and M. D.
Lukin,
 \prl{113}{243002}{2014}.





\bibitem{jun-ye-0}
K.-K. Ni, S. Ospelkaus, M. H. G. de Miranda, A. Pe'er, B. Neyenhuis, J. J. Zirbel, S. Kotochigova,
P. S. Julienne, D. S. Jin, and J. Ye, {\it Science} {\bf 322}, 231
(2008).

\bibitem{jun-ye-1}
M. H. G. de Miranda, A. Chotia, B. Neyenhuis, D. Wang, G. Qu\'em\'ener, S. Ospelkaus, J. L. Bohn, J. Ye, and D. S. Jin, 
{\it Nat. Phys.} {\bf 7} 502 (2011).

\bibitem{jun-ye-2}
A. Chotia, B. Neyenhuis, S. A. Moses, B. Yan, J. P. Covey, M. Foss-Feig, A. M. Rey, D. S. Jin, and J. Ye,
{\it Phys. Rev. Lett.} {\bf 108} 080405 (2012).


\bibitem{jun-ye-nature}
B. Yan, S. A. Moses, B. Gadway,	J. P. Covey, K. R. A. Hazzard, A. M. Rey, D. S. Jin and J. Ye,
{\it Nature} {\bf 501}, 521 (2013).


\bibitem{hazzard}
K. R. A. Hazzard, B. Gadway,  M. Foss-Feig, B. Yan, S. A. Moses, J. P. Covey, N. Y. Yao,  M. D. Lukin, J. Ye, D. S. Jin, and A. M. Rey, 
\prl{113}{195302}{2014}.



\bibitem{our-exciton-paper2}
F. Herrera, M. Litinskaya, and R. V. Krems, 
{\it Phys. Rev.} A {\bf 82}, 033428 (2010). 

\bibitem{ping-njp-paper}
P. Xiang, M. Litinskaya, E. A. Shapiro, and R. V. Krems, 
{\it New J. Phys.} {\bf 15}, 063015 (2013). 



\bibitem{DeMillePRL02}
D. DeMille, {\it Phys. Rev. Lett.} {\bf 88}, 067901 (2002). 


\bibitem{quantum-walk-1}
Y. Aharonov, L. Davidovich, and N. Zagury,
{\it Phys. Rev.} A {\bf 48}, 1687 (1993). 

\bibitem{quantum-walk-2}
S. R. Jackson, T. J. Khoo, and F. W. Strauch,
{\it Phys. Rev.} A {\bf 86}, 022335 (2012). 


\bibitem{biexcitons}
P. Xiang, M. Litinskaya and R. V. Krems, 
{\it Phys. Rev.} A {\bf 85}, 061401(R) (2012).  




\bibitem{our-polaron-paper1}
F. Herrera and R. V. Krems, {\it Phys. Rev.} A {\bf 84}, 051401(R) (2011).  



\bibitem{our-polaron-paper2}
F. Herrera, K. W. Madison, R. V. Krems, and M. Berciu, 
{\it Phys. Rev. Lett.} {\bf 110}, 223002 (2013). 


\bibitem{optical-lattices}
G. Grynberg and C. Robilliard, {\it Phys.
Rep.} {\bf 355}, 335 (2001).

\bibitem{note}
Quantum particles in 2D disordered potential are expected to undergo Anderson localization, provided the weak-localization correction is finite and negative. Depending on the miscrocopic details of the disorder potential, this may or may not be the case. Quoting from Ref. \cite{quote}: ``Scaling theory does not pretend to be an exact theory, there is thus real interest in knowing whether there is localization in 2 dimesions for specific systems''.

\bibitem{quote}
C. A. M\"uller and D. Delande, ``Disorder and interference: localization phenomena'', 
Chapter 9 in ``Les Houches 2009 - Session XCI: Ultracold Gases and Quantum Information'', edited by C. Miniatura et al. (Oxford University Press, 2011).


\bibitem{3D-anderson}
L. J. Roofl and J. L. Skinnerb, \jcp{89}{3279}{1989}. 

\bibitem{3D-note}
Note that the model considered here is an extreme case of a lattice with substitutional disorder, extensively studied in the literature, see, e.g., Ref. \cite{3D-anderson}. However, most studies consider models where particles are allowed to hop between nearest neighbours only and the conclusions of such studies must be modified to allow for long-range tunnelling to answer the questions raised in this work. 


\bibitem{agranovich}
V. M. Agranovich,
``{\it Excitations in Organic Solids}'', Oxford University Press, Oxford (2009). 


\bibitem{diffusion}
N. V. Prokof'ev and P. C. E. Stamp, {\it Phys. Rev.} A {\bf 74}, 020102(R) (2006).

\bibitem{diffusion-2}
C. D'Errico, M. Moratti, E. Lucioni, L. Tanzi, B. Deissler, M. Inguscio, G. Modugno, M.B. Plenio, and F. Caruso,
{\it New J. Phys.} {\bf 15}, 045007 (2013).

\bibitem{motional-narrowing}
Y. Sagi, R. Pugatch, I. Almog, N. Davidson, and M. Aizenman, 
\pra{83}{043821}{2011}. 

\bibitem{chaos-diffusion-1}
M. F. Shlesinger, G. M. Zaslavsky, and J. Klafter, {\it Nature (London)} {\bf 363}, 31 (1993).

\bibitem{chaos-diffusion-2}
M. Srednicki, \pre{50}{888}{1994}. 

\bibitem{vardi}
C. Khripkov, A. Vardi, and D. Cohen, arXiv:1406.6872.


\bibitem{dipolar}
T. Lahaye, C. Menotti, L. Santos, M. Lewenstein, and T. Pfau, 
{\it Rep. Prog. Phys.} {\bf 72}, 126401 (2009).

\bibitem{long-range-correlations}
G. Pupillo, A. Micheli, H. P. Buchler, and P. Zoller,
in ``{\it Cold Molecules: Theory, Experiment, Applications}'' (CRC Press, Boca Raton, 2009).

\bibitem{njp-review}
L. D. Carr, D. DeMille, R. V. Krems, and J. Ye,  {\it New. J. Phys.} {\bf 11}, 055049 (2009). 






\end{thebibliography}
\end{document}